\documentclass[sigconf]{acmart}

\usepackage{booktabs} % For formal tables

\setcopyright{rightsretained}
\usepackage{subfig}
\usepackage[draft]{pgf}
\usepackage{flushend}
\newcommand{\LM}[1]{}
\renewcommand{\LM}[1]{{\color{blue}[[LM: {#1}]]}}

\usepackage{xparse}

\newsavebox{\fminipagebox}
\NewDocumentEnvironment{fminipage}{m O{\fboxsep}}
{\par\kern#2\noindent\begin{lrbox}{\fminipagebox}
		\begin{minipage}{#1}\ignorespaces}
		{\end{minipage}\end{lrbox}%
	\makebox[#1]{%
		\kern\dimexpr-\fboxsep-\fboxrule\relax
		\fbox{\usebox{\fminipagebox}}%
		\kern\dimexpr-\fboxsep-\fboxrule\relax
	}\par\kern#2
}

\begin{document}
\title{Real-time Detection of Content Polluters in Partially Observable Twitter Networks}

\author{Mehwish Nasim}
%\authornote{Dr.~Trovato insisted his name be first.}
\affiliation{%
  \institution{School of Mathematical Sciences \\University of Adelaide}
  \city{Adelaide}
  \state{Australia}
}
\email{mehwish.nasim@adelaide.edu.au}

\author{Andrew Nguyen}
%\authornote{Dr.~Trovato insisted his name be first.}
\affiliation{%
	\institution{School of Mathematical Sciences \\University of Adelaide}
	\city{Adelaide}
	\state{Australia}
}
\email{andrew.nguyen03@adelaide.edu.au}

\author{Nick Lothian}
\authornote{Currently works at Tyto.ai}
\affiliation{%
	\institution{D2D CRC}
	\city{Adelaide}
	\state{Australia}
}
\email{info@d2dcrc.com.au}

\author{Robert Cope}
%\authornote{Dr.~Trovato insisted his name be first.}
\affiliation{%
	\institution{School of Mathematical Sciences \\University of Adelaide}
	\city{Adelaide}
	\state{Australia}
}
\email{robert.cope@adelaide.edu.au}

\author{Lewis Mitchell}
\authornote{D2D CRC Stream Lead}

\affiliation{%
	\institution{School of Mathematical Sciences \\University of Adelaide}
	\city{Adelaide}
	\state{Australia}
}
\email{lewis.mitchell@adelaide.edu.au}

\fancyhead{}

\begin{abstract}
Content polluters, or bots that hijack a conversation for political or advertising purposes are a known problem for event prediction, election forecasting and when distinguishing real news from fake news in social media data. 
Identifying this type of bot is particularly challenging,
with state-of-the-art methods utilising large volumes of network data as features for machine learning models. 
Such datasets are generally not readily available in typical applications which stream social media data for real-time event prediction. 
In this work we develop a methodology to detect content polluters in social media datasets that are streamed in real-time. 
Applying our method to the problem of civil unrest event prediction in Australia, 
we identify content polluters from individual tweets,
without collecting social network or historical data from individual accounts. 
We identify some peculiar characteristics of these bots in our dataset and propose metrics for identification of such accounts. 
We then pose some research questions around this type of bot detection, including: how good Twitter is at detecting content polluters and how well state-of-the-art methods perform in detecting bots in our dataset. 

\end{abstract}


\copyrightyear{2018}
\acmYear{2018}
\setcopyright{iw3c2w3}
\acmConference[WWW '18 Companion]{The 2018 Web Conference Companion}{April 23--27, 2018}{Lyon, France}
\acmBooktitle{WWW '18 Companion: The 2018 Web Conference Companion, April 23--27, 2018, Lyon, France}
\acmPrice{}
\acmDOI{10.1145/3184558.3191574}
\acmISBN{978-1-4503-5640-4/18/04}

\begin{CCSXML}
	<ccs2012>
	<concept>
	<concept_id>10002951.10003227.10003233.10010519</concept_id>
	<concept_desc>Information systems~Social networking sites</concept_desc>
	<concept_significance>500</concept_significance>
	</concept>
	<concept>
	<concept_id>10002978.10003022.10003027</concept_id>
	<concept_desc>Security and privacy~Social network security and privacy</concept_desc>
	<concept_significance>300</concept_significance>
	</concept>
	</ccs2012>
\end{CCSXML}

\ccsdesc[500]{Information systems~Social networking sites}
\ccsdesc[300]{Security and privacy~Social network security and privacy}

\keywords{Civil unrest, Social bots, Content polluters, Missing links, Twitter}

\maketitle

\section{Introduction}
%darpa bot challenge

\subsection{Motivation}

%Do hypothesis testing!
%
%Write the comments of study participants!
%
%``\textit{Without news to feed it, the biggest story starves}"- \\Emlyn Williams.

Bots and content polluters in online social media affect the socio-political state of the world, from meddling in elections \cite{buzzfeed, bessi2016social, ferrara2017disinformation} to influencing US veterans \cite{junkoxford}. In late September 2017, Twitter admitted to Congress that it had found 200 Russian accounts that overlapped with Facebook accounts which were used to sway Americans and create divisions during the elections held in 2016 \cite{buzzfeed}.  
Of course, some bots are useful as well, for instance accounts that will tweet alerts to people about natural disasters. The problem arises when they try to influence people or spread misinformation. 
The importance of detecting bots in online social media has produced an active research area on this topic \cite{kayes2017privacy, cresci2017paradigm}.

State-of-the-art methods for bot detection use historical patterns of behaviour and a rich feature set including textual, temporal, and social network features, to distinguish automated bots from real human users \cite{Varol2017}.
However, for real-time application using large streamed datasets, such methods can be prohibitive due to the sheer volume, velocity, and incompleteness of data samples.
In this work we develop a new method to detect one particular type of social bot -- content polluters -- in streamed microblog datasets such as Twitter. 
Content polluters are bots that attempt to subvert a genuine discussion by hijacking it for political or advertising purposes.
As we will show, these bots are a major concern for applications such as real-time event prediction, such as social unrest, from social media datasets.

\subsection{Problem context}

Social unrest prediction is a growing concern for governments worldwide. This is evidenced by DARPA's Open Source Intelligence program, which produced numerous methods to predict the occurrence of future population-level events such as civil unrest, political crises, election outcomes and disease outbreaks \cite{doyle2014forecasting,muthiah2015planned,ramakrishnan2014beating,saraf2016embers}. It has been observed that social events are either preceded or followed by changes in population-level communication behaviour, consumption and movement. 
A large fraction of population-level changes are implicitly reflected in online data such as blogs, online social networks, financial markets, or search queries. Some of these data sources have been shown to effectively detect population-level events in real time. Methods have been developed for predicting such events by fusing publicly available data from multiple sources.
There exists a plethora of research focused on social media-based forecasting models, suggesting that features from micro-blogs such as Twitter can predict and detect population-level events \cite{ramakrishnan2014beating}. Once one develops a ``gold standard" (ground truth) record of known events (e.g. election results, or protests occurring) models can be trained using open source data to make predictions. 
A significant challenge for such models is noise reduction through filtering ``fake news'', removing misclassified or irrelevant tweets, or mitigating the effects of missing data. 
This is of particular concern, as the changing limits on accessing social media data remains a major challenge for researchers \cite{nasim2016investigating}. 
Access to data through APIs and third parties can be inconsistent, incomplete, and corrupted by noise in the form of bots. 
Where bots are influencing people through fake social media accounts, they also act as \textit{content polluters} on social media sites \cite{suarez2016influence}. 
According to the Digital Forensics Research Lab (DFRL), ``They can make a group of six people look like a group of 46,000 people."

The main goal of our work was finding out content polluters in a dataset comprising tweets related to Australian social unrest events in real time, without access to complete profile information of the users. 
Due to rate limits on the public API and the high cost of accessing data, we were restricted to using only streamed tweets satisfying certain criteria. 
While the actual event prediction algorithm is not the primary concern of this paper, further detail can be found in Osborne \emph{et al.}~\cite{grant}.

% While we focus on Australian event prediction here, the challenges also hold for event prediction, anywhere in the world.
%"There’s no surefire way to spot whether an account is a bot. But there are a number of different ways to look at Twitter accounts — mostly to determine whether their behaviour resembles that of a human — that may help us gain a better understanding of the universe of automated user accounts. In short, bots tend to tweet more and in bursts, have automated or copied features in their profiles, and retweet a lot."

\begin{table*}[ht]
	\caption{Data statistics}
	\label{tab:stats}
	\centering
	\begin{tabular}{ l |l | l |l|l |l}
		\hline
		
		Parameters &  Adelaide & Brisbane &  Melbourne & Perth & Sydney\\
		\hline
		\hline
		Number of tweets  & $14087$& $5913$ & $23720$& $8421$ & $31568$\\
		Number of unique users   & $12039$& $3466$ & $14611$& $6215$ & $14515$\\
		%Number of unique hashtags & 1 & a & 1 & 1&1 &1& 1 & 1\\
		Number of unique URLs  & $548$& $233$ &$762$& $456$ & $844$\\
		Average number of followers (in degree)   & $8812$& $9624$ & $6733$& $5409$ & $6052$\\
		Average number friends (out degree)   & $1223$& $1736$ & $1517$& $1643$ & $1860$\\
		Number of verified accounts   & $293$& $432$ & $840$ & $209$ & $412$\\
		\hline
	\end{tabular}
\end{table*}

\subsection{Related Work}

% In this section we summarize the literature on the detection of social bots, sometimes referred to as automated accounts/fake accounts or content polluters. 
A social bot is a computer algorithm that automatically produces content and interacts with humans on social media, trying to emulate and possibly alter their behaviour \cite{socialbots}. 
Social bots inhabit social media platforms, and online social networks are inundated by millions of bots exhibiting increasingly sophisticated, human-like behaviour. In the coming years a proliferation of social media bots is expected as advertisers, criminals, politicians, governments, terrorists, and other organizations attempt to influence populations \cite{darpa2016}. This introduces dimensions for social bots, including social network characteristics, temporal activity, diffusion patterns, and sentiment expression \cite{socialbots}. 

Ghost \emph{et al.}~\cite{ghosh2012understanding} conducted an analysis on the follower/followee links acquired by over 40,000 spammer accounts suspended by Twitter. They showed that penalizing users for connecting to spammers can be effective because it would de-incentivize users from linking with other users in order to gain influence. 
Yang \emph{et al.}~\cite{yang2014uncovering} found that bot accounts in online social networks connect to each other by chance and integrate into the social network just like normal users.  
Network information along with content has been shown to detect spam in online social networks \cite{hu2014online}. 
While researchers were proposing various bot-detection models, Lee \emph{et al.}~\cite{lee2014will} identified and engaged strangers on social media to effectively propagate information/misinformation. 
They proposed a model to leverage peoples' social behaviour (online interactions) and users' wait times for retweeting. 

Social bots evolve over time, making them resilient against standard bot detection approaches \cite{cresci2017paradigm}. They are apt at changing discussion topics and posting activities \cite{wu2017detecting}. Researchers have proposed complex models,
such as those based on interaction graphs of suspicious accounts \cite{hu2014online,yang2013empirical, han2013efficient, keller2017manipulate}. 
An adversary often controls multiple social bots known as a \emph{sybil}. 
One strategy to detect such accounts relies on investigating social graph structure, on the assumption that sybil accounts link to a small number of legitimate users \cite{cao2012aiding}. Behavioural patterns and sentiments analysis have also been used for bot detection \cite{dickerson2014using}. Such patterns can easily be encoded in features, thus machine learning techniques can be used to distinguish bot-like from human-like behaviour. Previous work uses network-based features or content analysis for bot detection, along with indicators such as temporal activity, retweets, and crowd sourcing \cite{truthy, wang2015making}. Such efforts require substantial network knowledge or the ability to quickly query an API for a complete history of social media postings by suspected bots. However, real-time applications, such as streaming messages based on keywords or geographic locations, render this impractical. A major challenge therefore is developing methodologies to detect and remove bots based on partial information, message histories, and network knowledge, in real time.

In this work we detect bots from individual tweets downloaded for predicting social unrest in Australian cities.  
%\textit{Beat the news} is an initiative that seeks to predict civil unrest events in Australia\footnote{\textit{Beat the news} initiative is a partnership between industry and Australian academia that seeks to develop technology that will automatically and accurately predict the occurrence of future population-level events such as civil unrest.}. 
% It is the first study of its type to predict civil unrest via streamed social media at this scale in Australia. 
Given filters on keywords and geographic location of events (such as protests, rallies, civil disturbances) collected in real time, it leaves a small but informative dataset for prediction. Predictions are generated in real time by analysing data from online social media platforms such as Twitter and validated against hand-labeled ``Gold standard records'' (\textit{GSR}) \cite{grant}. 
The GSR is created by the news analysts; after going through a validation and cleaning process this data is ready to be used as the ground truth. If Twitter data is contaminated with social bots, it can greatly degrade prediction models. 
It is therefore imperative to develop techniques for detecting and removing social bots for real-time data streams.

\textbf{Contributions:} Our scientific contributions are as follows:

\begin{enumerate}
	
	\item We develop a method to identify social bots in data using only partial information about the user and their tweet history, in real time. 
	
	\item We present a new dataset of hand-labelled bots and legitimate records, and use it to validate our method\footnote{Data can be accessed on http://maths.adelaide.edu.au/mehwish.nasim/}. 
	
	\item We pose a set of research questions for evaluating whether Twitter users, Twitter, or existing state-of-the-art bot detection methods could detect bots in our dataset or not.
	
\end{enumerate}

\subsection{Dataset}

Our dataset consists of timestamped tweets from 1 January 2015 till 31 December 2016 from $5$ major capital cities in Australia. Tweets identify one of the following locations: `Australia', `Adelaide', `Brisbane', `Melbourne', `Perth', or `Sydney'. The data are targeted at studying civil unrest and intends to capture ways in which people express opinions and organize marches, rallies, peaceful/violent protests etc., within Australia. Such events aim to draw attention toward an issue e.g., infrastructure, taxes, immigration laws etc. Australia has a population of about $24.5$ million people and, like in many developed countries, predicting civil unrest events is of interest to law enforcement agencies, government bodies, media and academia. 
Notwithstanding this fact, the literature is devoid of exploratory studies conducted on this population for real-time prediction of civil unrest events. The basic statistics about protest-related tweets in our dataset are reported in Table \ref{tab:stats}.

Note that the dataset was devoid of information on the alters (followers/friends of egos), except for the total count of alters (numbers of followers and friends).

%\textbf{Structural Properties:}
%The dataset consists of about $20$ million tweets. 
%However, the sample of interest are only the protest-related tweets. Basic data statistics about protest related tweets are reported in Table \ref{tab:stats}. We carried out an exploratory analysis on our dataset consisting of tweets, basic profile information of the \textit{egos}, and number of followers and friends without downloading the followers/friends mutual friendship graphs, as in most large scale real-time analysis this is prohibitive due to API rate limits. 

% Theoretically, it was possible to download more data, such as follower friends network or apply further filters on the data, but we decided to stick to the data that was being used by other project members for event prediction. 
\begin{figure}[t]
	\centering
	\subfloat[Two purple nodes at the right side that are loosely connected to the core, are bots. They have tweeted together frequently and their individual frequency to tweet is low as compared to other nodes in the graph, however the dyadic (pairwise) frequency is higher.]{\includegraphics[width=0.8\columnwidth]{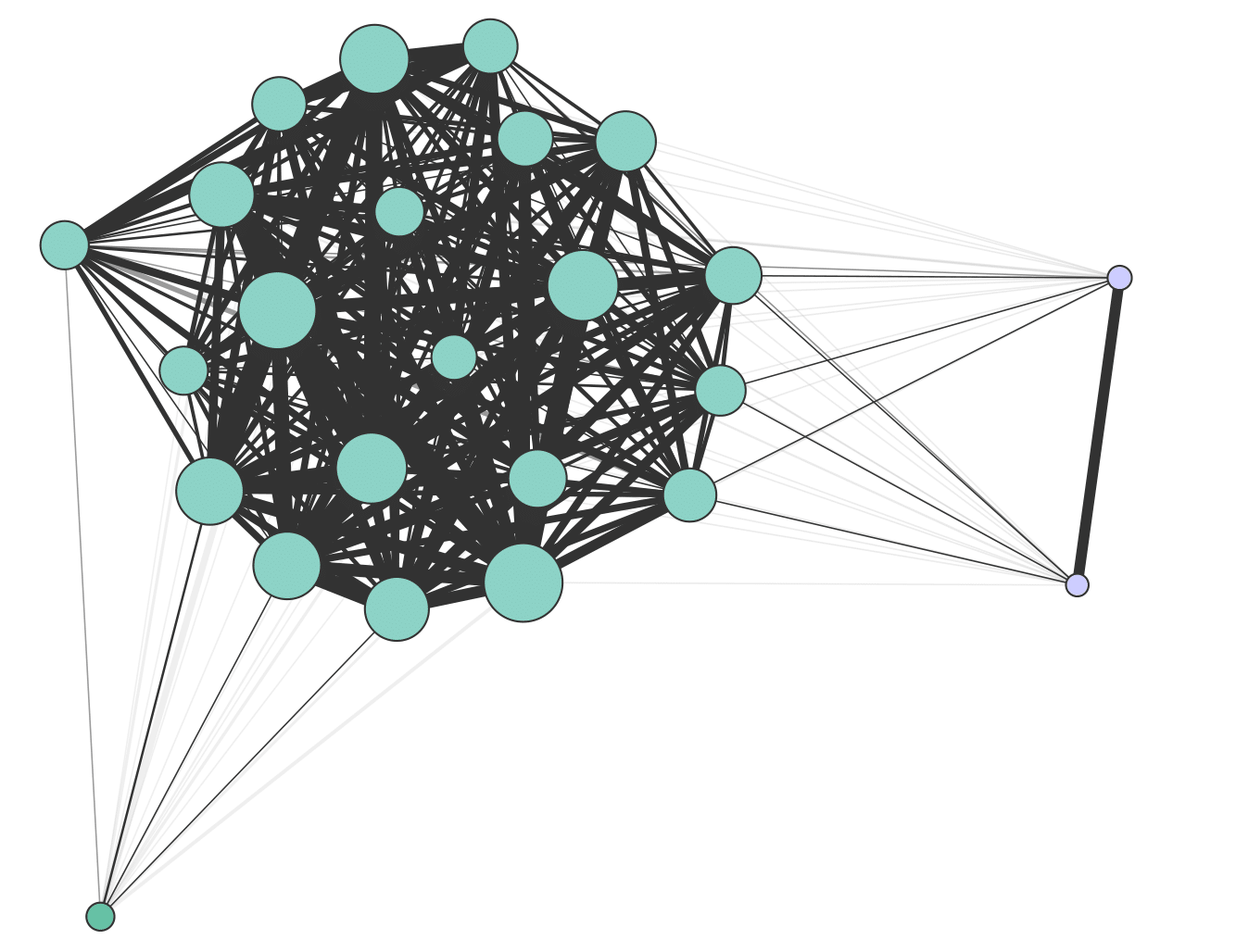} \label{fig:botsoneday} }\qquad 
	\subfloat[Two densely connected components in the tweets graph.]{\includegraphics[width=0.9\columnwidth]{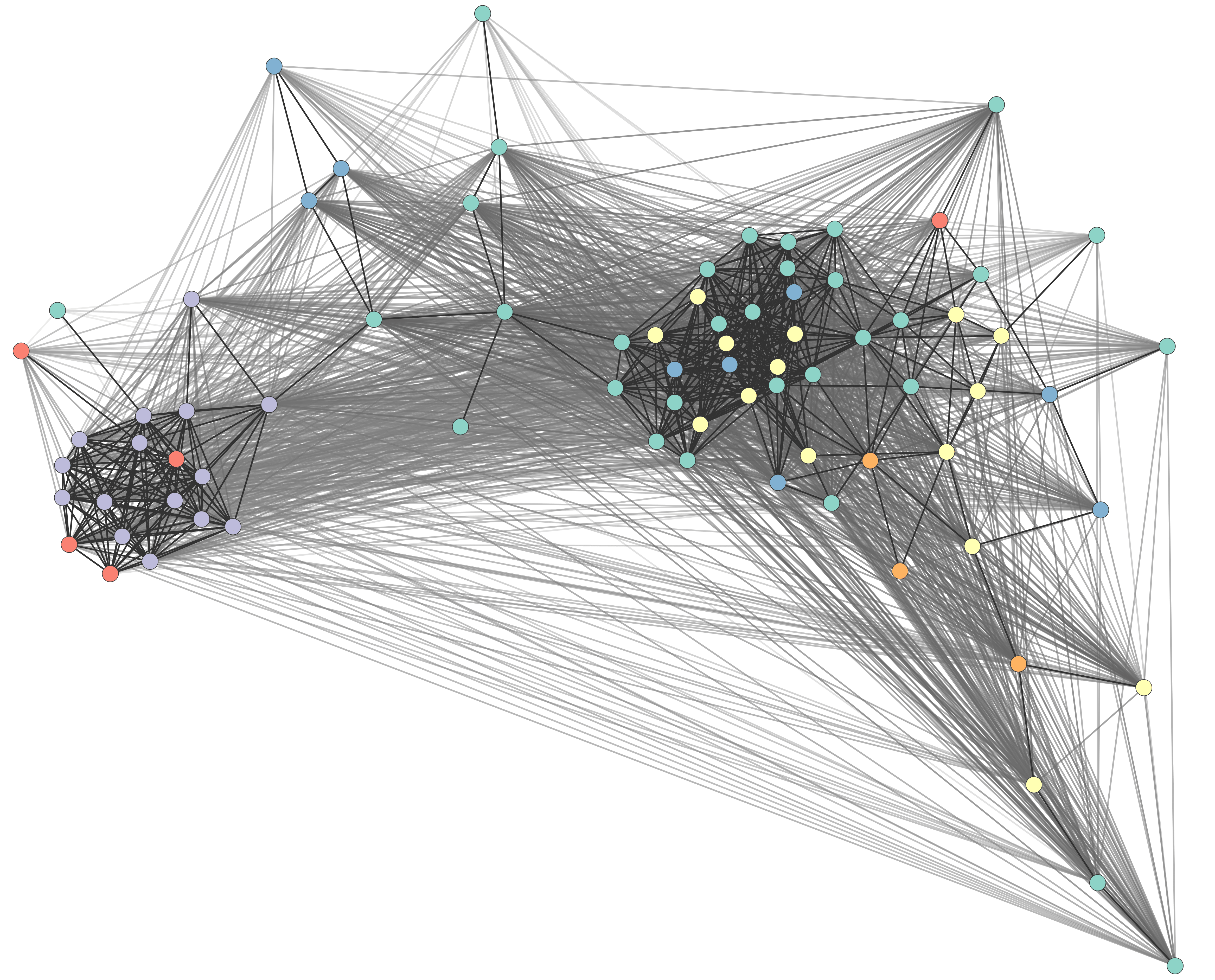} \label{fig:events} }\qquad
%	\subfloat[Vertex induced subgraph from the tweets network. Bots are the purple nodes. Green nodes represent false positives. Orange nodes are not bots but are also not relevant for event predictions, since these users were not geographically located in Australia and were tweeting about Victoria in the UK]{\includegraphics[width=0.9\columnwidth]{botsincore} \label{fig:botsincore}}
	\caption{Graphs containing bots and legitimate users from the Melbourne events network.}
	\label{fig:egoandpersonal}
\end{figure}

\begin{figure}[thb]
	\centering
	\includegraphics[width=0.7\columnwidth]{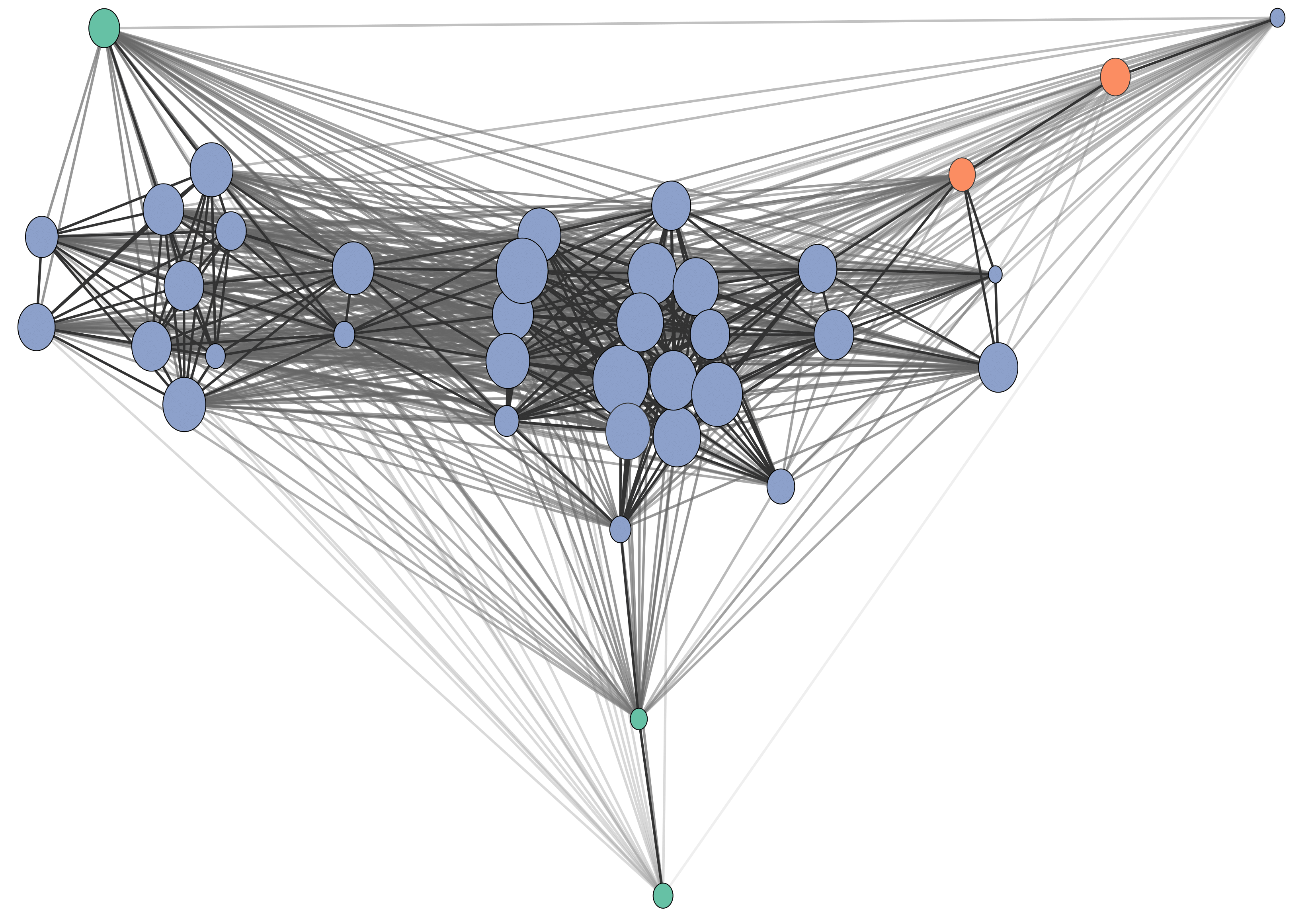}
	\caption{Graph containing bots and legitimate users from the Melbourne events network.}
	\label{fig:botsincore}
\end{figure}

\section{Detecting content polluters}

We investigate two characteristics of tweets i.e., temporal information and message diversity in a tweet. 

%Tweets focused on protests in the five major Australian cities. The tweets consisted of keywords such as, \textit{protest} and \textit{rally}.
\begin{figure*}[ht]
	\centering
	\subfloat[Legitimate (Gini = 0.8, $R^2 = 0.98$)]{\includegraphics[width=0.65\columnwidth]{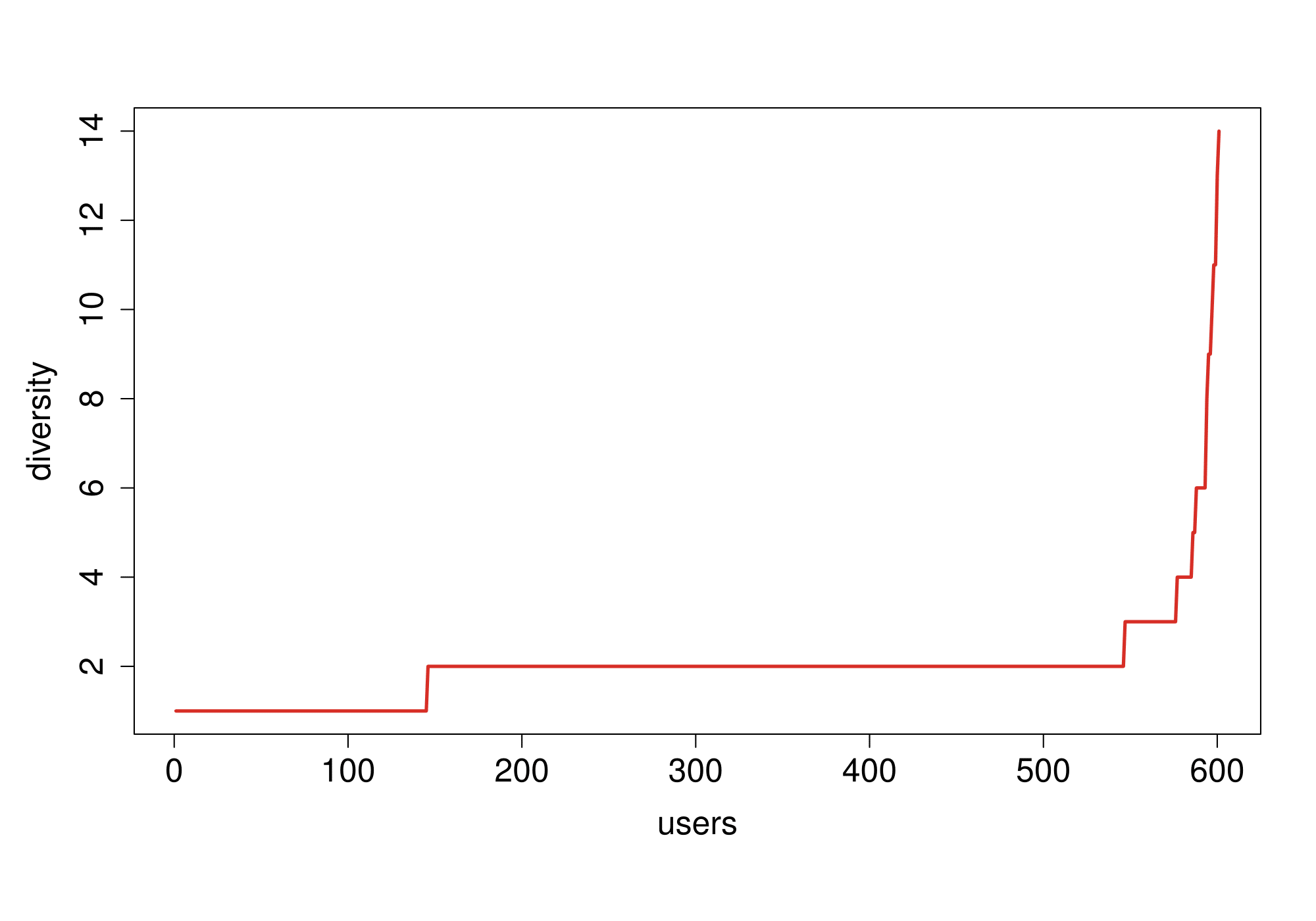} \label{fig:youtube} }\quad
	\subfloat[Bots (Gini = 0.32, $R^2 = 0$)]{\includegraphics[width=0.65\columnwidth]{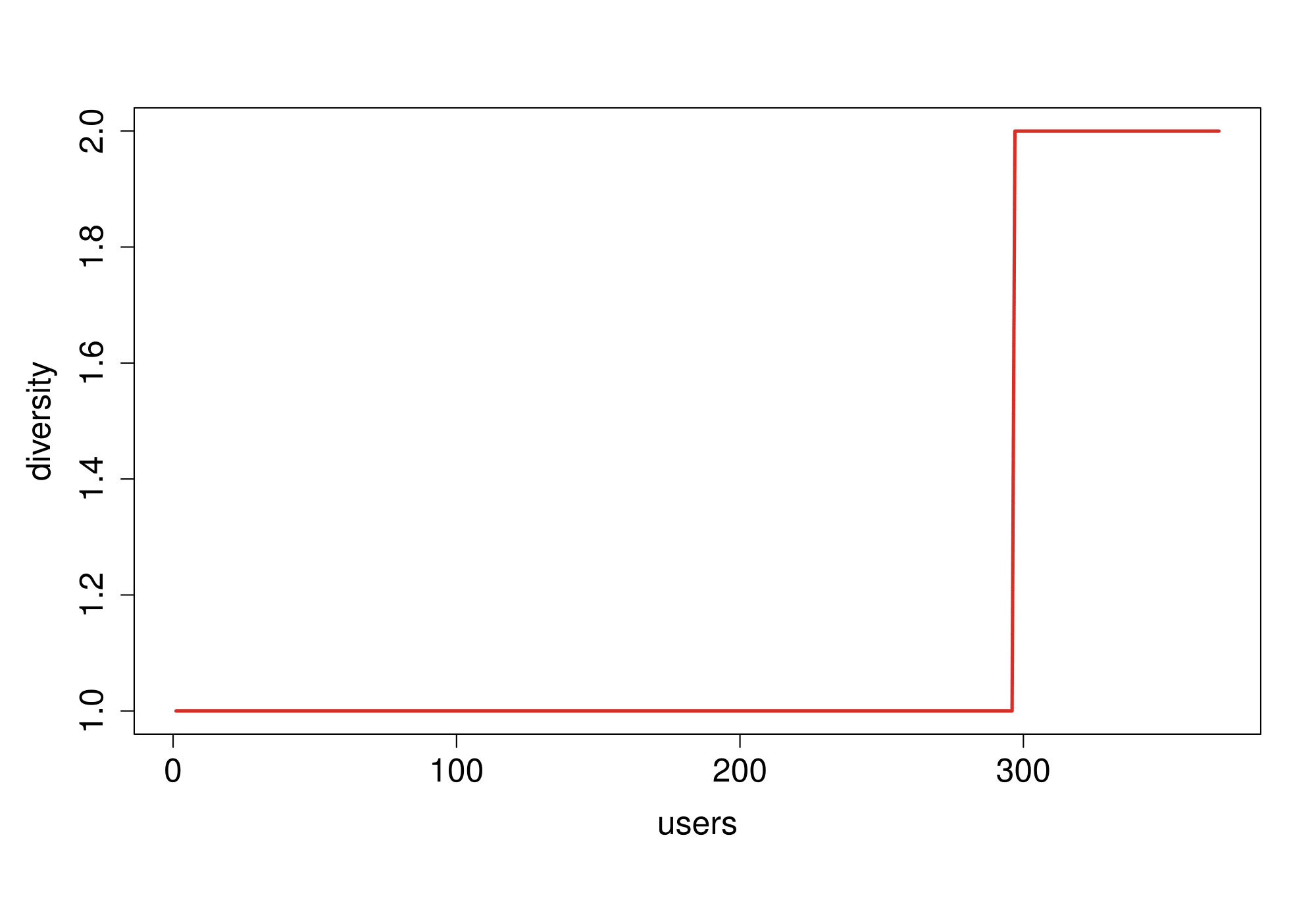} \label{fig:moja}}\quad
	\subfloat[Bots (Gini = 0, $R^2 = 0$)]{\includegraphics[width=0.65\columnwidth]{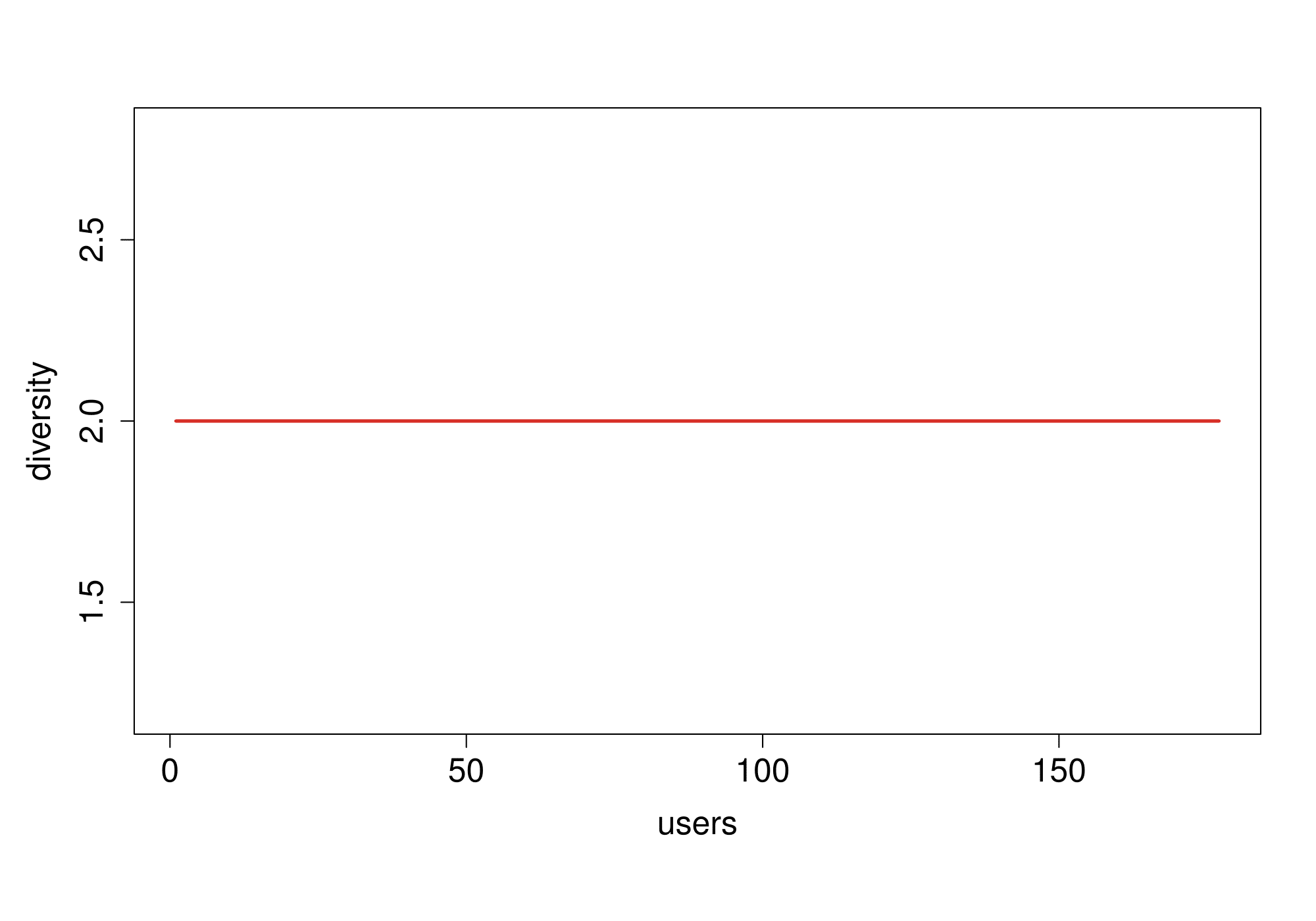} \label{fig:dm}}
	\caption{Message diversity measured through 3 URLs for bots and genuine users.}
	\label{fig:urlsResults}
\end{figure*}

\textbf{Temporal Patterns:} 
In the first step we were interested in 1). users who tweet frequently, 2). pairs of users who tweet on the same day using the desired keywords.
Since no information about the network of individual users is available, we cannot construct a follower-friend network graph. Instead, we construct a two mode user-event network. For all the events in the data we connect two users if they have tweeted on the same event day. We represent this problem in graph theoretic terms as follows:

Let $G$ be a bipartite graph of users and events. Let $U$ be the set of users and let $V$ be the set of events.
Let $ u, v \in U$ and let $i, j \in V$.  For any $i \in V$ if $N(u) \cap N(v) \neq \{ \}$
then  $(u,v) \in E $ in the one-mode projection of the bipartite graph. The \textit{neighbourhood} $N(v)$ of a vertex $v \in U$ is the set of vertices that are adjacent to $v$. 
The resulting projection is an undirected loopless multigraph. If the edge set $E$ contains the same
edge several times, then $E$ is a multiset. If an edge occurs several times in $E$, the
copies of that edge are called parallel edges. Graphs that have parallel edges are also called multigraphs.

\begin{figure}[thb]
	\centering
	{\includegraphics[width= 8cm]{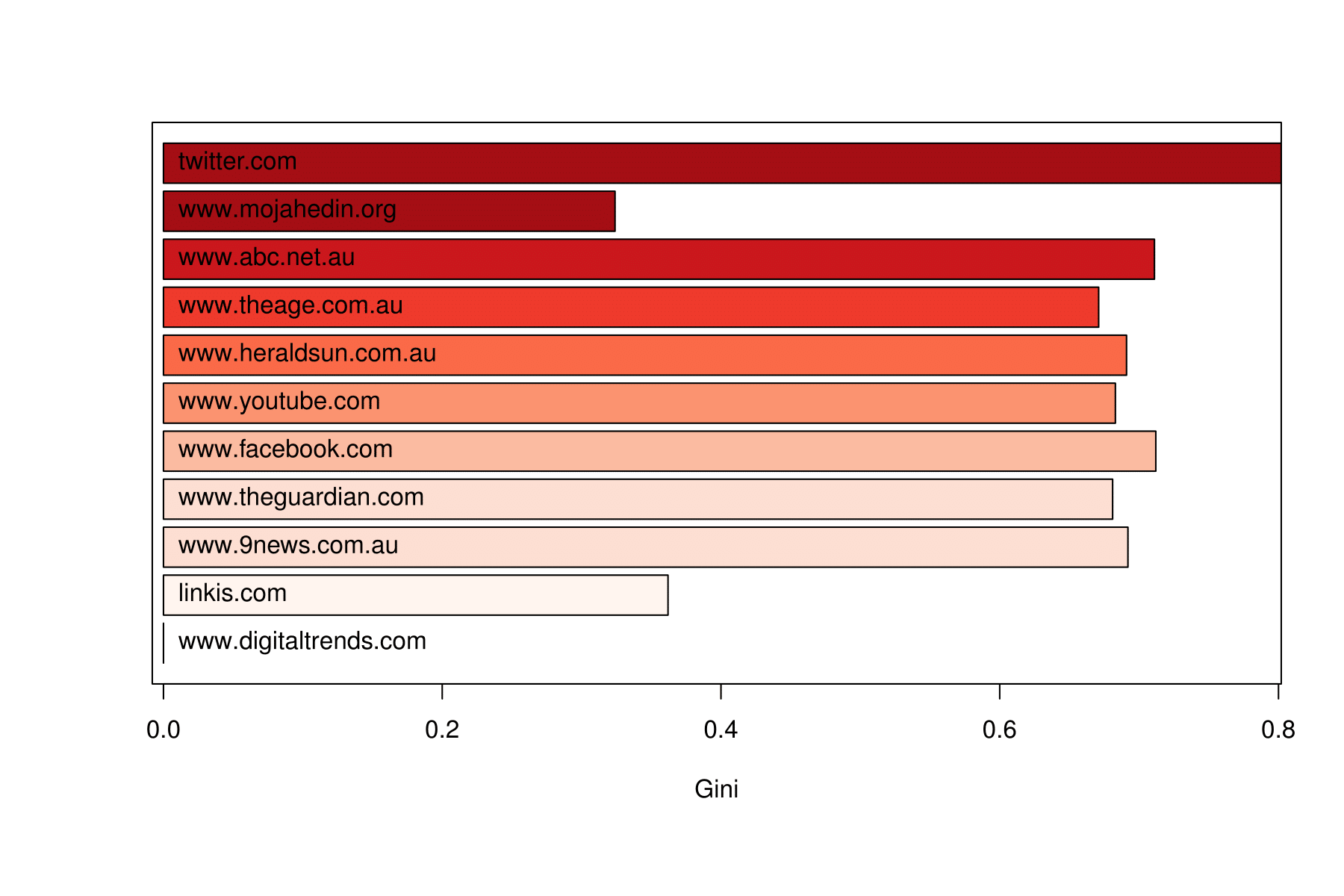}  }
	\caption{Gini score for ten URLs. High Gini coefficient indicates a legitimate URL. The three URLs with the lowest Gini coefficients were being tweeted by content-polluting bots.}
	\label{fig:urlsPlot}
\end{figure}

Similar to other social networks such as friendship networks, event networks are a result of complex sociological processes with a multitude of relations. When such relations are conflated into a dense network, the visualization often resembles a ``hairball''. Various approaches to declutter drawings of such networks exist in the literature. We use the recent \textit{backbone layout} approach for network visualization \cite{nocaj2014untangling}, which accounts for strong ties (or multiplicity of edges) and uses the union of all maximum spanning trees as a sparsifier to ensure a connected subgraph. In Figure \ref{fig:events}, the thickness of edges represents how often a pair of nodes tweet on the same 'event day'\footnote{Event day was confirmed from the \textit{GSR}.} whereas, the size of the nodes indicates the individual frequency of tweets by a user\footnote{Networks visualizations are created in \textit{visone} (http://www.visone.info/).}. We noticed that bots tweeted together frequently. Their individual frequency to tweet is low as compared to other nodes in the graph, however the dyadic (pairwise) frequency is higher. 
For instance, the two purple nodes on the right have tweeted together frequently, in Figure \ref{fig:botsoneday}. Their individual frequency to tweet is low as compared to other nodes in the graph, however the dyadic (pairwise) frequency is higher. These two nodes are weakly connected to the core. Upon checking their complete profiles, the users were found to be political bots. This motivated us to further explore the tweets-graph. 

The core of the network (green nodes) were found to be news channels and popular blogs in Australia, such as \textit{MelbLiveNews}, \textit{newsonaust}, \textit{7NewsMelbourne} and \textit{LoversMelbourne} to name a few. Media accounts are likely to report population-level events on the day of the events, thus they form a strongly-connected core of the events network graph. 

We then clustered all tweets in a similar manner to construct a graph where two users have an edge between them if they have tweeted on the same day, irrespective of whether there was an event that day or not. We used the Louvain Method for clustering the network \cite{blondel2008fast}, based on the concept of \emph{modularity}. Optimizing the modularity results in the best possible grouping of nodes in a given network. We then found two strongly-connected components in the graph: 1. News channels, and 2. Bots. We analysed the strongly-connected vertex-induced subgraphs from the network. One such component for the city of \textit{Melbourne} is shown in Figure \ref{fig:botsincore}, which is a strongly-connected component from Figure \ref{fig:events}. Bots are the purple nodes (validated by manual inspection of profiles). Green nodes represent false positives. Orange nodes are not bots but are also not relevant for predictions, since these users were not geographically located in Australia and were tweeting about Victoria in the UK.

%\textbf{Diversity score:}

\textbf{Message diversity:} We computed the diversity in the tweets based upon mentions of URLs and hashtags. We selected the top most tweeted URLs, $\{K\}$ ($|K|=20$), and then filtered out the users ($\bar{U} \subseteq U$) who mentioned those URLs. The motivation for this approach is that an event prediction model should be resilient against bot-URLs that are infrequently mentioned in the tweets, so these will not greatly impact the prediction accuracy. 
We then computed the following three measures for each of the remaining users: i). total number of tweets containing any URL(s), $u^{all}_i$, ii). number of tweets mentioning URL $k \in K$, $u^k_i$ and iii). diversity score i.e., the difference between the two measures, $u^d_i = u^{all}_i - u^k_i$.

We then plot the diversity score distribution for every $u^k \in \bar{U}$, for every URL $k \in K$. This immediately provides some relevant insights about the behaviour of content polluters: 
Figure \ref{fig:youtube} shows a legitimate URL (i.e., linked to by legitimate users), whereas, Figures \ref{fig:moja} and \ref{fig:dm} show bot-URLs (i.e., URLs linked to by bots). 
Users who tweet these URLs are classified as \textit{potential bots}. 
The figures show that the diversity of users linking to legitimate URLs is generally far greater than those linking to bot-URLs.
The temporal patterns of bot-URL mentions and those which are being tweeted at regular intervals indicated that these users were indeed bots. 

We measure the extent of diversity in two ways:
\begin{enumerate}
	\item \textit{The Gini} coefficient ($G \in \mathcal{R}$, G=[0,1]):
	\begin{equation}
	G = \frac{\sum_{i=1}^{n} \sum_{j=1}^n | u^d_i - u^d_j|}{2n \sum_{i=1}^n u^d_i},
	\end{equation}
	where $n$ is the number of users tweeting a particular URL.
	
	The Gini coefficient $G$ describes the relative amount of inequality in the distribution of diversity: $G = 0$ indicates complete equality while $G = 1$ indicates complete inequality. 
	A high $G$ suggests coordination among the observations. 
	The Gini coefficient does not measure absolute inequality and the interpretation can vary from situation to situation. Legitimate accounts such as news channels, newspapers, and famous activists are likely to tweet legitimate and diverse URLs, thus the Gini coefficient for legitimate URLs is high as compared to illegitimate URLs. The Gini coefficient for a sample of ten URLs is  shown in Figure \ref{fig:urlsPlot}.
	
	\item \textit{Rank-size Rule:} We observed that only a fraction of URLs are mentioned very frequently in the tweets and very large number of URLs barely find their way in more than a single tweet. It is interesting to note that cities and their rank also follow a similar distribution; this pattern is generally known as the \textit{rank-size rule} \cite{ranksize}. This has also been observed in various studies on calling behaviour of users \cite{bentley2015composition}\cite{bergman2012you} \cite{nasim2016mobilehci}.
	\end {enumerate}
	
	We fit a curve on every user versus URL-diversity graph and measure the coefficient of determination $R^2$. Values close to zero indicate that the model explains little of the variability of the response data around its mean. For legitimate URLs, we obtained values close to 1 (Figure \ref{fig:urlsResults}). 
	
	Recently, Gilani \emph{et al.}~\cite{gilaniclassification} evaluated the characteristics of automated versus human accounts by looking at complete tweet histories. They initially hypothesized that bots tweet a number of different URLs, however in the actual data they found that humans may also post a number of URLs. Conversely, in this work we looked at most frequently posted URLs and then for each URL we analysed how diverse the users' tweets are who are tweeting that URL.

\begin{figure*}[th]
			\centering
			
			\subfloat[Account creation time]{\includegraphics[width=0.9\columnwidth]{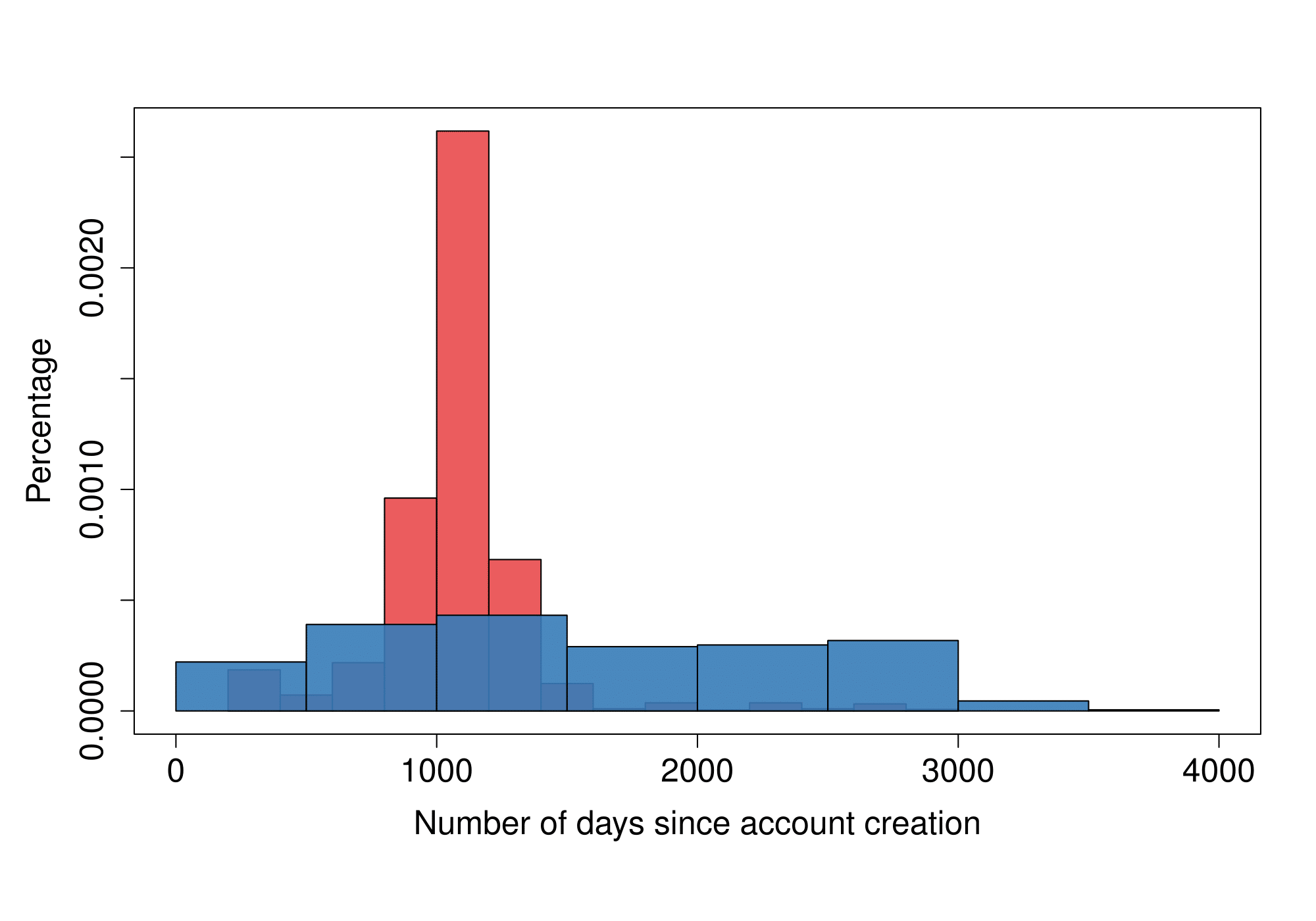} \label{fig:accountcreation} }\quad
			\subfloat[Length of Twitter account name]{\includegraphics[width=0.9\columnwidth]{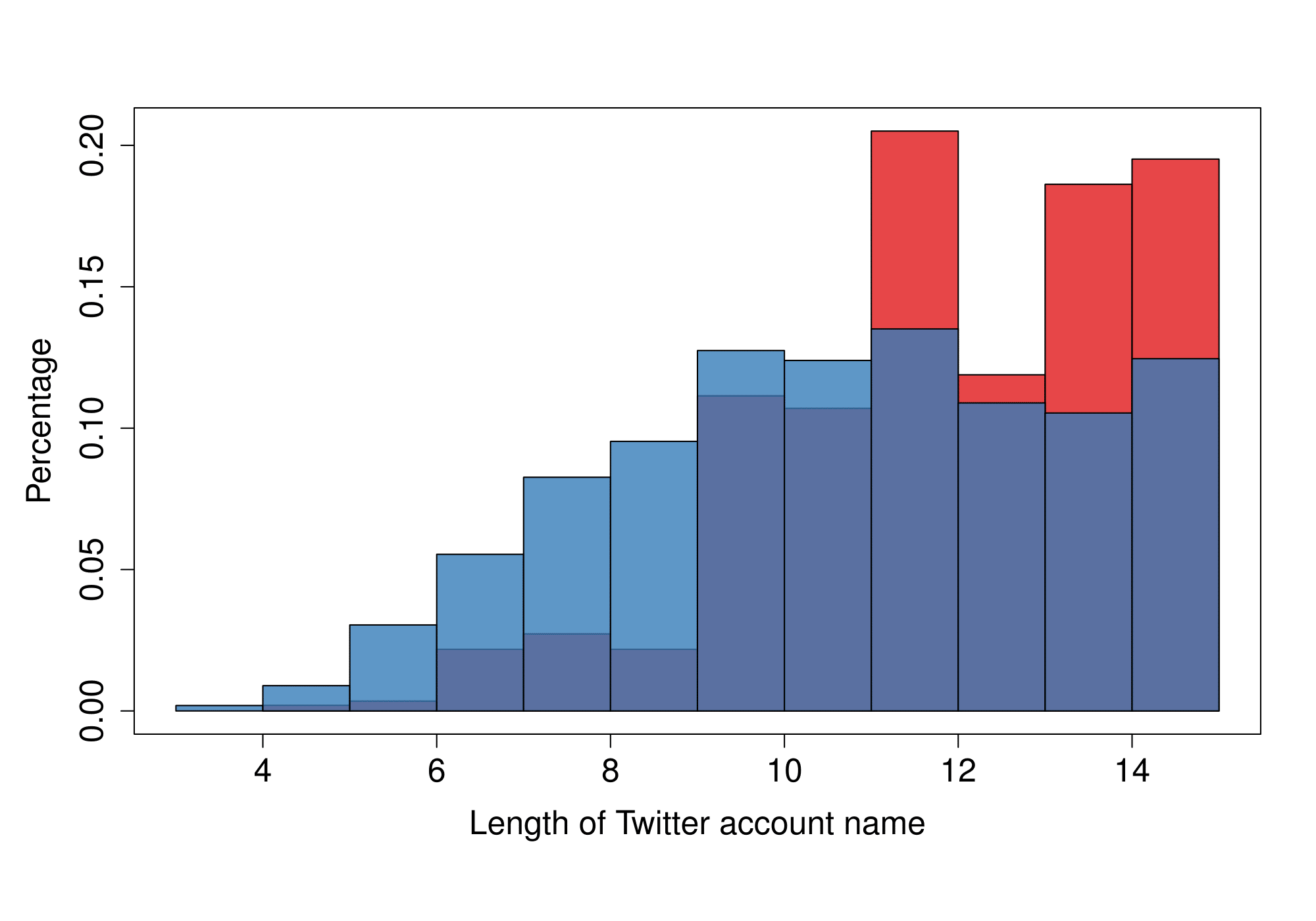} \label{fig:namelength}}
			\caption{Characteristics of users. Bots(red) versus Legitimate users(blue)}
			\label{fig:botscharacteristics}
		\end{figure*}

	We detected $849$ bots in the data using message diversity on URLs, which we call \emph{content polluters}. These content polluters contributed about $7\%$ of tweets in the data. We computed some statistics on content polluters versus legitimate users, shown in Figure \ref{fig:botscharacteristics}. 
	In \cite{socialbots}, authors argued that social bots tend to have recent accounts with long names. 
	However, we did not find a significant difference in our data between content polluters and regular users. 
	The average account age of content polluters accounts was $2.9$ years as compared to legitimate users which was $4.2$ years.
	This difference was significant ($p < 0.01$).
	%\todo{Is the difference actually significant? Can you compute a p-value here please?}
	This suggests that these particular type of bot accounts are relatively old and have remained (potentially) undetected by Twitter. 
	The length of Twitter names for bots had on average $11$ characters as compared to non-bots that had $12$ characters. 
	None of the bots had \textit{verified} Twitter accounts. A total of $109$ political bot accounts were created on 20 February 2014 with only $12$ unique names, a strong indication of being a bot network. 
	We also found several digital media bot accounts. Such accounts aim at becoming famous by attracting followers \cite{boshmaf2011socialbot}. 
	A set of such accounts was created on 30 March 2016. 
	This set consisted of $8$ accounts with an average friend count of $4099$ and follower count of $1112$.
	
	We also explored the dataset from \cite{lee2011seven} using our algorithm. The dataset contains more than 600k tweets. The Gini coefficient for each dataset (bots and non-bots) was around 0.5, hence we remain inconclusive. 
	The data set from Gilani \emph{et al.}~\cite{gilaniclassification} only consisted of the number of URLs each user mentioned, therefore it was not possible to check the relative frequency of any particular URL.
	We argue that the nature of content polluting bots makes them difficult to distinguish in traditional bot-detection datasets.
	This motivates our research questions below and the creation of a new human-validated content-pollution dataset in the next section.

	\section{Creating a content-polluting bot dataset}\label{sec:researchquestions}
	
	Given the peculiarities in the bot accounts that we found in our analysis, we move on to some pertinent research questions. 
	
	\subsection{Do humans succeed in detecting content polluters?}

	We conducted a user study to hand-label a set of Twitter accounts that contained equal number of content polluters (from our list obtained in the previous section) and legitimate accounts. We asked three independent hand-labellers to create the dataset. Users were first shown several examples of content polluters as well as of legitimate accounts. All three participants were well versed with using Twitter. All participants found it very difficult to assess non-English accounts even with automatic translation.
	
	The participants recorded the following comments:
	
	\begin{center}
		\begin{fminipage}{0.95\columnwidth} 
			\textit{Participant 1}
			
			Domain Knowledge: Advance Twitter User
			
			Comments: \textit{``What I'm struggling with is that, the user doesn't actually initiates a suspicious tweet. He simply retweets a whole bunch of content polluting tweets"}.\\
			
			Strategy: 
			\begin{itemize}
				\item If user has tweeted or retweeted from well known news spam sites then mark as bot.
				\item Otherwise look through pattern of tweets, if very spamy tweet behaviour, for example highly consistent frequency of tweeting behaviour and tweets are from a single source then mark as bot.
				\item See if they regular mention and interact with other twitter users which indicates a good sign for a regular account.
				\item Look at profile details and follower and followees ratio to distinguish if it appears like a regular account or a bot.
			\end{itemize}
			
		\end{fminipage}
	\end{center}
	
	\begin{center}
		\begin{fminipage}{0.95\columnwidth} 
			\textit{Participant 2}
			
			Domain Knowledge: Twitter User/Domain Expert
			
			Comments: \textit{``This was a really hard task. It is not at all clear what the difference is between a bot and a human. This is much slower than labelling individual tweets.}"\\
			Strategy: 
			\begin{itemize}
				\item Look at the twitter account. If the user has tweeted well-known news spam URLs/services (@convoy, dv.it, 360WISE, mojahedin.org) then mark it as a bot.
				\item If not, scroll through the account. If I can find some original content (e.g., comments on a retweet) then mark it as a legitimate account.
				\end{itemize}
			
		\end{fminipage}
	\end{center}
	
	% user 2 cont
		\begin{center}
			\begin{fminipage}{0.95\columnwidth} 
				\textit{Participant 2 continued..}
				\begin{itemize}
					\item If it always retweets from a single or only a few accounts then I mark it as a bot.
					\item Otherwise it comes down to judgement. This includes things like looking at avatar icons - if all the followers seem strangely similar (all Anime figures for example, or all faded pictures), or if it always uses the same non-twitter link shortener then I mark it as a bot.
					\item Then there were a set of accounts who all posted the same content. I only noticed this after a while, so I probably only caught some of them.
					\item There are a set of Markov-chain-like accounts (e.g. 1240541203). It can be difficult to distinguish them because of the messy, non-standard language typical of many Twitter users, and the limited amount of text to work from.
				\end{itemize}
				
			\end{fminipage}
		\end{center}

	% user 3
	\begin{center}
		\begin{fminipage}{0.95\columnwidth} 
			\textit{Participant 3}
			
			Domain Knowledge: Everyday Twitter User
			
			Comments: \textit{`` I think there are lots of accounts that are part automatic and potentially part human (e.g., with the annoying ``I've gained/lost n followers" tweets), which seems like a challenge. I tried to work out if these were actually human."}
			
			Strategy: 
			
			\begin{itemize}
				\item In each case, I would skim through the tweets. The presence of coherent original content without a URL suggested that the account was likely human.
				\item Tweets from recognisable spam sources (e.g. 360WISE) suggested the account was a bot.
				\item Overwhelming consistency across tweets suggested a bot (i.e., every tweet with exactly the same text formatting, or using the same URL shortener), except when this was associated with what appeared to be a carefully curated account for a business.
				\item For the rest, I looked at profile information, follower ratios, and the broader content of the tweets and made the best judgement I could.
			\end{itemize}
			
		\end{fminipage}
	\end{center}

		\begin{figure}[hbt]
			\centering
			{\includegraphics[width=0.9\columnwidth]{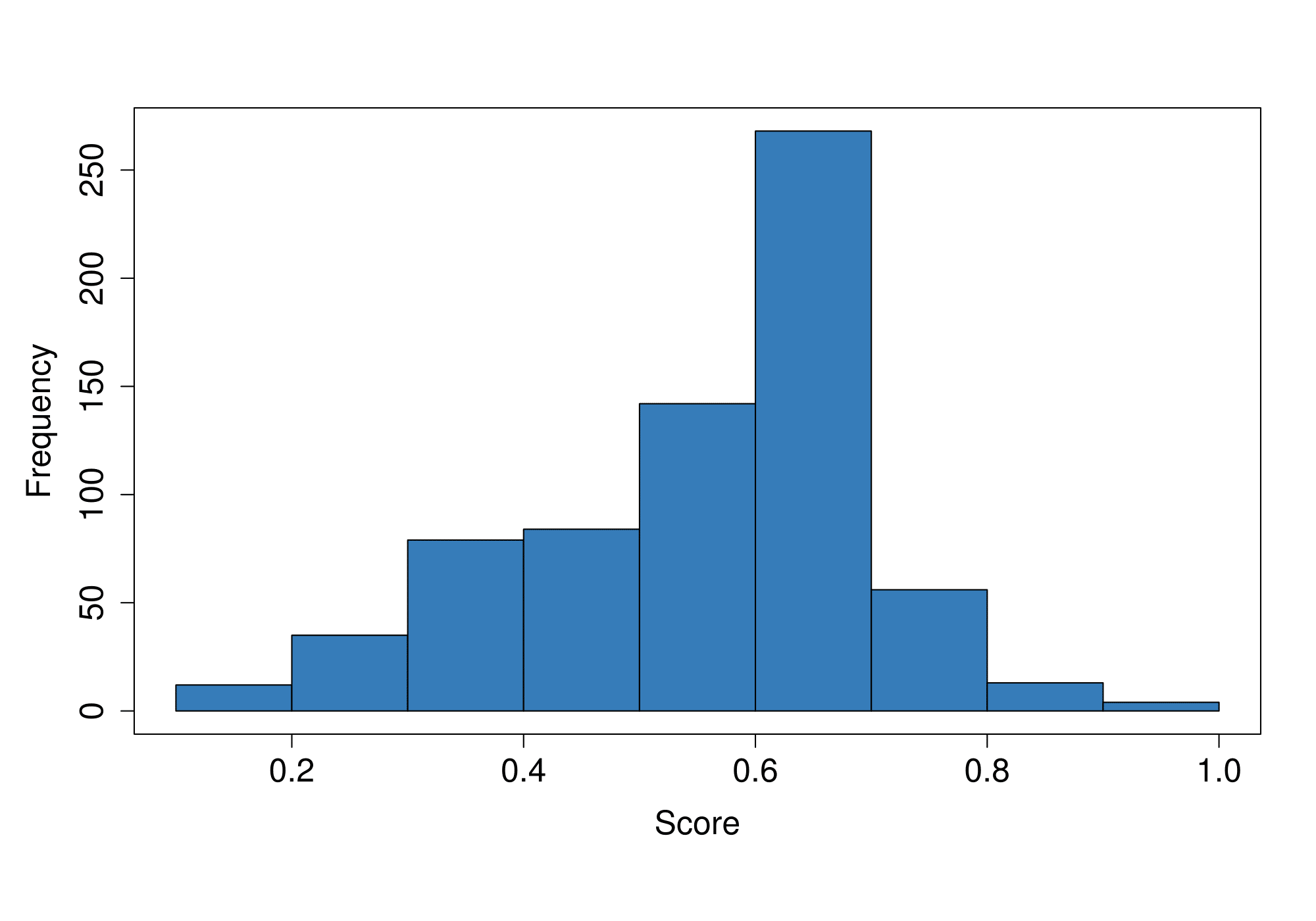}  }
			\caption{Results obtained from Truthy on the complete list of bots. Please note that the results were obtained after removing the 153 accounts that were suspended by Twitter.}
			\label{fig:truthy}
		\end{figure}
	
	We constructed a labelled dataset from this collection by recording where 2 out of the 3 hand-labellers agreed on a classification.
	For our content polluter algorithm, we observed that the proportion of observed correct prediction (for both the classes) on this hand-labelled dataset was 0.57. 
	
	We test the following hypothesis:
	$\mathbf{H1_a}$: Our method was able to find bot/non-bot accounts with greater than $50\%$ accuracy. 	Hence, the null hypothesis is:	
	$\mathbf{H_0}$: Our method randomly labelled the bot/non-bot accounts. 
	
	After applying a $t$ test we reject the null hypothesis at a significance level of $\alpha = 0.05$ ($p = 0.00029$).
	
	\subsection{How efficient is Twitter in detecting social bots?}
	
	Twitter continuously searches for suspicious accounts, and accounts that are found to be malicious may be deleted. In this experiment we studied how many bot accounts Twitter suspended from our detected bots list. The dataset that we analysed is from 2015/2016 but we conducted this experiment in April 2017, giving us a comprehensive set of accounts determined by Twitter to be bots. We used the Twitter API for this experiment. 
	Given a query for a specific account, the Twitter API returns an error message if the account is suspended by Twitter or deleted by the user. It the error code~$63$ is returned then it means Twitter has suspended the account, whereas, error code~$50$ means the user deleted the account. 
	For active accounts, metadata information about the account is returned. 	Upon querying the Twitter API, we found that Twitter had suspended $153$ accounts out of the $849$ content polluters that we have detected. 
	
	\begin{table*}[ht]
		\caption{Performance summary of our method versus Truthy. According to Truthy, $65\%$ of the true positives in the user study were likely to be a bot, whereas, $21\%$ of false positives also had a greater than $0.5$ probability to be a bot.}
		\label{tab:summary}
		\centering
		\begin{tabular}{ l | l |l|l }
			\hline
			
			& &Truthy($\mu$) & Truthy ($\sigma$)  \\
			\hline
			\hline
			%Bots detected using diversity score  & 696 & 137  &   & &153  \\
			%Bots in user study (detected by diversity score)  & 276& &   &  &14 \\
			True Positives reported by user study & 65\% & 0.556  & 0.159   \\
			False Positives reported by user study & 21\% & 0.392 & 0.131  \\
			
			\hline
		\end{tabular}
	\end{table*}
	
	\subsection{How efficient other methods are for bot detection?}
	
	%\textbf{For instance results from truthy}
	
	We also tested the performance of state-of-the-art bot detection system called \textit{Truthy} \cite{truthy}, also known as BotOrNot? \cite{Varol2017}.
	It is a publicly 
	available API service developed by the Indiana University at Bloomington in May 2014 to evaluate the similarity of a Twitter account with the known characteristics of social bots. It uses the complete profile of a user to determine how likely the user account is to be a bot.
	% To the best of our knowledge it is the only publicly-available social bot detection system.
	BotOrNot? employs a supervised machine-learning classifier that exploits more than 1000 features from the Twitter account under investigation. Features are derived based on network information and tweeting behaviour. The authors state that despite the fact that the service is specifically designed for detection of bots, the performance against evolved spambots might be worse than that was reported in the paper. We queried this service against our genuine set of bot accounts. 
	Truthy displays scores against each account. Higher scores mean more bot-like accounts. 
	% The two users in Figure \ref{fig:botsoneday} were given a score of $0.43$ and $0.25$. 
	Figure \ref{fig:truthy} shows the overall performance of Truthy against our list of bots. The mean score was $0.55$ with a standard error of $0.14$. Table \ref{tab:summary} shows the performance summary.
	We remark that for the task of detecting content polluters our method performs comparably to Truthy,
	using only the information of URL diversity at the sampled-tweet level.
	We reiterate that we detected content polluter accounts using message diversity since we did not have access to complete account information, whereas Truthy exploited features obtained from the complete user profile and network. 
	However, the aim of Truthy is much different from what we are trying to achieve. 
	We utilize \textbf{single tweets} with users' metadata to filter out bots in real-time for event prediction.

	\section{Discussion}
	
	In this work we  discovered social bots in protest-related tweets, using a stream of sampled Twitter data. 
	Unlike previous bot-detection studies, our dataset was devoid of network information and detailed tweeting history. 
	We showed the efficacy of a start-of-the-art Twitter-bot detection technique using complete profile and network information on our dataset. 
	Further, we analysed the capabilities of Twitter and naive users in distinguishing bots from legitimate users. Twitter continuously looks for malicious accounts, which are deleted. 
	However, this process may be very slow and a number of accounts remain undetected \cite{bastos2017brexit}. It is difficult to track users who delete their tweets, since Twitter does not provide access to deleted tweets, or tweets older than 30 days.
	We show that even in the presence of complete network information, existing methods are not apt at detecting content-polluting bots. We argue that for real-time Twitter streams where it is difficult to obtain detailed profile information because of constraints on time and scalability, a cost-effective way is to compute the message diversity of tweets for each user. A low diversity might indicate suspicious accounts. The most challenging aspect of this work is to validate results since user perceptions are not always correct, and standard bot detection methods are very much prone to misclassification despite using complete twitter account information \cite{cresci2017paradigm, gilani2017bots,gilaniclassification}. 
	Results from our user study indicate that our method agrees with the participants on most accounts that are legitimate. However, there is some difference in opinion for the bot accounts since it largely involves human perception of what a bot or a content polluter could be. For instance, participant-$3$ indicated that they did not consider an account to be a bot when this was associated with what appeared to be a carefully curated account for a business. However, when we looked into the original tweets, certain users used hashtags such as `$\#melbourne$', while promoting their business that had nothing to do with the city Melbourne. Participants also indicated that some accounts seemed to be part automated and potentially part human. Even advanced Twitter users found distinguishing between bots and legitimate users a challenging task. 
	Deletion of tweets is a major issue for traditional bot detection methods. In the case of US elections, a recent news article says, ``Twitter is either unable or unwilling to retrieve a substantial amount of tweets from bots and fake users spreading disinformation. Those users, which have been tied to Russia, have since deleted those tweets" \cite{Mashable}. In the absence of malicious tweets any validation method is prone to failure.  
	
	\textbf{Performance Improvement:} In February 2017 we used our content-polluters detection methodology in order to improve the performance of a social media-based predictive model \cite{grant}. 
	Users from a large Australian law enforcement agency provided positive feedback on improvements in the predictions. We noted that the model was no longer erroneously predicting events related to `escorts', which improved model performance noticeably. Further, removal of bots also removed $18$ non-interesting events in the month of February, related to lottery ticket sales.

	\section{Conclusion and Future Work}
	
	We found that content polluters in this dataset often timed their tweets together. By analysing the temporal patterns one could infer the presence of bot accounts. However, we also noticed that tweets from news channels were also temporally correlated. Using only temporal methods could lead to misclassification of legitimate accounts. We also found that bots used a small set of URLs in their tweets, therefore by finding out the most frequently-used URLs and computing their relative usage in tweets from the all the unique users in the dataset, one could successfully detect content polluters. Our analysis leads us to believe that conventional machine learning methods may require a number of features and may not be apt at correctly identifying bots. The bots that we detected in our dataset helped to remove noise in the data and significantly improved the performance of  prediction models. In future we aim to:
	\begin{enumerate}
		\item Analyse non-protest related tweets for detecting bots, and utilise other available relations such as user-event relations, temporal relations, social interactions etc.
		
		\item Characterize bots into various categories and explore whether some bots could even be useful for civil unrest prediction.
		
		\item Conduct more user studies with a larger number of participants, in order to further understand the characteristics of content polluters.
	\end{enumerate}

\section{ACKNOWLEDGEMENTS}
The authors acknowledge financial support from Data to Decisions CRC. MN and LM also acknowledge support from the ARC Centre of Excellence for Mathematical and Statistical Frontiers (ACEMS).

\bibliographystyle{ACM-Reference-Format}
	\bibliography{references}


\begin{thebibliography}{40}

%%% the Web address.

\ifx \showCODEN    \undefined \def \showCODEN     #1{\unskip}     \fi
\ifx \showDOI      \undefined \def \showDOI       #1{#1}\fi
\ifx \showISBNx    \undefined \def \showISBNx     #1{\unskip}     \fi
\ifx \showISBNxiii \undefined \def \showISBNxiii  #1{\unskip}     \fi
\ifx \showISSN     \undefined \def \showISSN      #1{\unskip}     \fi
\ifx \showLCCN     \undefined \def \showLCCN      #1{\unskip}     \fi
\ifx \shownote     \undefined \def \shownote      #1{#1}          \fi
\ifx \showarticletitle \undefined \def \showarticletitle #1{#1}   \fi
\ifx \showURL      \undefined \def \showURL       {\relax}        \fi
% invisible to TeX
\providecommand\bibfield[2]{#2}
\providecommand\bibinfo[2]{#2}
\providecommand\natexlab[1]{#1}
\providecommand\showeprint[2][]{arXiv:#2}

\bibitem[\protect\citeauthoryear{Bastos and Mercea}{Bastos and Mercea}{2017}]%
        {bastos2017brexit}
\bibfield{author}{\bibinfo{person}{Marco~T Bastos} {and} \bibinfo{person}{Dan
  Mercea}.} \bibinfo{year}{2017}\natexlab{}.
\newblock \showarticletitle{The Brexit Botnet and User-Generated Hyperpartisan
  News}.
\newblock \bibinfo{journal}{\emph{Social Science Computer Review}}
  (\bibinfo{year}{2017}), \bibinfo{pages}{0894439317734157}.
\newblock


\bibitem[\protect\citeauthoryear{Bentley and Chen}{Bentley and Chen}{2015}]%
        {bentley2015composition}
\bibfield{author}{\bibinfo{person}{Frank Bentley} {and}
  \bibinfo{person}{Ying-Yu Chen}.} \bibinfo{year}{2015}\natexlab{}.
\newblock \showarticletitle{The Composition and Use of Modern Mobile
  Phonebooks}. In \bibinfo{booktitle}{\emph{Proceedings of the 33rd Annual ACM
  Conference on Human Factors in Computing Systems}}. ACM,
  \bibinfo{pages}{2749--2758}.
\newblock


\bibitem[\protect\citeauthoryear{Bergman, Komninos, Liarokapis, and
  Clarke}{Bergman et~al\mbox{.}}{2012}]%
        {bergman2012you}
\bibfield{author}{\bibinfo{person}{Ofer Bergman}, \bibinfo{person}{Andreas
  Komninos}, \bibinfo{person}{Dimitrios Liarokapis}, {and}
  \bibinfo{person}{James Clarke}.} \bibinfo{year}{2012}\natexlab{}.
\newblock \showarticletitle{You never call: Demoting unused contacts on mobile
  phones using DMTR}.
\newblock \bibinfo{journal}{\emph{Personal and Ubiquitous Computing}}
  \bibinfo{volume}{16}, \bibinfo{number}{6} (\bibinfo{year}{2012}),
  \bibinfo{pages}{757--766}.
\newblock


\bibitem[\protect\citeauthoryear{Bessi and Ferrara}{Bessi and Ferrara}{2016}]%
        {bessi2016social}
\bibfield{author}{\bibinfo{person}{Alessandro Bessi} {and}
  \bibinfo{person}{Emilio Ferrara}.} \bibinfo{year}{2016}\natexlab{}.
\newblock \showarticletitle{Social bots distort the 2016 US Presidential
  election online discussion}.
\newblock  (\bibinfo{year}{2016}).
\newblock


\bibitem[\protect\citeauthoryear{Blondel, Guillaume, Lambiotte, and
  Lefebvre}{Blondel et~al\mbox{.}}{2008}]%
        {blondel2008fast}
\bibfield{author}{\bibinfo{person}{Vincent~D Blondel},
  \bibinfo{person}{Jean-Loup Guillaume}, \bibinfo{person}{Renaud Lambiotte},
  {and} \bibinfo{person}{Etienne Lefebvre}.} \bibinfo{year}{2008}\natexlab{}.
\newblock \showarticletitle{Fast unfolding of communities in large networks}.
\newblock \bibinfo{journal}{\emph{Journal of statistical mechanics: theory and
  experiment}} \bibinfo{volume}{2008}, \bibinfo{number}{10}
  (\bibinfo{year}{2008}), \bibinfo{pages}{P10008}.
\newblock


\bibitem[\protect\citeauthoryear{Boshmaf, Muslukhov, Beznosov, and
  Ripeanu}{Boshmaf et~al\mbox{.}}{2011}]%
        {boshmaf2011socialbot}
\bibfield{author}{\bibinfo{person}{Yazan Boshmaf}, \bibinfo{person}{Ildar
  Muslukhov}, \bibinfo{person}{Konstantin Beznosov}, {and}
  \bibinfo{person}{Matei Ripeanu}.} \bibinfo{year}{2011}\natexlab{}.
\newblock \showarticletitle{The socialbot network: when bots socialize for fame
  and money}. In \bibinfo{booktitle}{\emph{Proceedings of the 27th annual
  computer security applications conference}}. ACM, \bibinfo{pages}{93--102}.
\newblock


\bibitem[\protect\citeauthoryear{Cao, Sirivianos, Yang, and Pregueiro}{Cao
  et~al\mbox{.}}{2012}]%
        {cao2012aiding}
\bibfield{author}{\bibinfo{person}{Qiang Cao}, \bibinfo{person}{Michael
  Sirivianos}, \bibinfo{person}{Xiaowei Yang}, {and} \bibinfo{person}{Tiago
  Pregueiro}.} \bibinfo{year}{2012}\natexlab{}.
\newblock \showarticletitle{Aiding the detection of fake accounts in large
  scale social online services}. In \bibinfo{booktitle}{\emph{Proceedings of
  the 9th USENIX conference on Networked Systems Design and Implementation}}.
  USENIX Association, \bibinfo{pages}{15--15}.
\newblock


\bibitem[\protect\citeauthoryear{Chin}{Chin}{2017}]%
        {Mashable}
\bibfield{author}{\bibinfo{person}{Monica Chin}.}
  \bibinfo{year}{2017}\natexlab{}.
\newblock \bibinfo{title}{Report: Twitter deleted tweets related to the Russian
  investigation}.
\newblock   (\bibinfo{year}{2017}).
\newblock
\urldef\tempurl%
\url{http://mashable.com/2017/10/13/twitter-deleted-russian-tweets/#CIbGh7BglkqS}
\showURL{%
\tempurl}


\bibitem[\protect\citeauthoryear{Cresci, Di~Pietro, Petrocchi, Spognardi, and
  Tesconi}{Cresci et~al\mbox{.}}{2017}]%
        {cresci2017paradigm}
\bibfield{author}{\bibinfo{person}{Stefano Cresci}, \bibinfo{person}{Roberto
  Di~Pietro}, \bibinfo{person}{Marinella Petrocchi}, \bibinfo{person}{Angelo
  Spognardi}, {and} \bibinfo{person}{Maurizio Tesconi}.}
  \bibinfo{year}{2017}\natexlab{}.
\newblock \showarticletitle{The paradigm-shift of social spambots: Evidence,
  theories, and tools for the arms race}. In
  \bibinfo{booktitle}{\emph{Proceedings of the 26th International Conference on
  World Wide Web Companion}}. International World Wide Web Conferences Steering
  Committee, \bibinfo{pages}{963--972}.
\newblock


\bibitem[\protect\citeauthoryear{Davis, Varol, Ferrara, Flammini, and
  Menczer}{Davis et~al\mbox{.}}{2016}]%
        {truthy}
\bibfield{author}{\bibinfo{person}{Clayton~Allen Davis}, \bibinfo{person}{Onur
  Varol}, \bibinfo{person}{Emilio Ferrara}, \bibinfo{person}{Alessandro
  Flammini}, {and} \bibinfo{person}{Filippo Menczer}.}
  \bibinfo{year}{2016}\natexlab{}.
\newblock \showarticletitle{BotOrNot: A system to evaluate social bots}. In
  \bibinfo{booktitle}{\emph{Proceedings of the 25th International Conference
  Companion on World Wide Web}}. International World Wide Web Conferences
  Steering Committee, \bibinfo{pages}{273--274}.
\newblock


\bibitem[\protect\citeauthoryear{Dickerson, Kagan, and Subrahmanian}{Dickerson
  et~al\mbox{.}}{2014}]%
        {dickerson2014using}
\bibfield{author}{\bibinfo{person}{John~P Dickerson}, \bibinfo{person}{Vadim
  Kagan}, {and} \bibinfo{person}{VS Subrahmanian}.}
  \bibinfo{year}{2014}\natexlab{}.
\newblock \showarticletitle{Using sentiment to detect bots on Twitter: Are
  humans more opinionated than bots?}. In \bibinfo{booktitle}{\emph{Advances in
  Social Networks Analysis and Mining (ASONAM), 2014 IEEE/ACM International
  Conference on}}. IEEE, \bibinfo{pages}{620--627}.
\newblock


\bibitem[\protect\citeauthoryear{Doyle, Katz, Summers, Ackermann, Zavorin, Lim,
  Muthiah, Butler, Self, Zhao, et~al\mbox{.}}{Doyle et~al\mbox{.}}{2014}]%
        {doyle2014forecasting}
\bibfield{author}{\bibinfo{person}{Andy Doyle}, \bibinfo{person}{Graham Katz},
  \bibinfo{person}{Kristen Summers}, \bibinfo{person}{Chris Ackermann},
  \bibinfo{person}{Ilya Zavorin}, \bibinfo{person}{Zunsik Lim},
  \bibinfo{person}{Sathappan Muthiah}, \bibinfo{person}{Patrick Butler},
  \bibinfo{person}{Nathan Self}, \bibinfo{person}{Liang Zhao}, {et~al\mbox{.}}}
  \bibinfo{year}{2014}\natexlab{}.
\newblock \showarticletitle{Forecasting significant societal events using the
  Embers streaming predictive analytics system}.
\newblock \bibinfo{journal}{\emph{Big data}} \bibinfo{volume}{2},
  \bibinfo{number}{4} (\bibinfo{year}{2014}), \bibinfo{pages}{185--195}.
\newblock


\bibitem[\protect\citeauthoryear{Ferrara}{Ferrara}{2017}]%
        {ferrara2017disinformation}
\bibfield{author}{\bibinfo{person}{Emilio Ferrara}.}
  \bibinfo{year}{2017}\natexlab{}.
\newblock \showarticletitle{Disinformation and social bot operations in the run
  up to the 2017 French presidential election}.
\newblock  (\bibinfo{year}{2017}).
\newblock


\bibitem[\protect\citeauthoryear{Ferrara, Varol, Davis, Menczer, and
  Flammini}{Ferrara et~al\mbox{.}}{2016}]%
        {socialbots}
\bibfield{author}{\bibinfo{person}{Emilio Ferrara}, \bibinfo{person}{Onur
  Varol}, \bibinfo{person}{Clayton Davis}, \bibinfo{person}{Filippo Menczer},
  {and} \bibinfo{person}{Alessandro Flammini}.}
  \bibinfo{year}{2016}\natexlab{}.
\newblock \showarticletitle{The rise of social bots}.
\newblock \bibinfo{journal}{\emph{Commun. ACM}} \bibinfo{volume}{59},
  \bibinfo{number}{7} (\bibinfo{year}{2016}), \bibinfo{pages}{96--104}.
\newblock


\bibitem[\protect\citeauthoryear{Gallacher and Kelly}{Gallacher and Kelly}{[n.
  d.]}]%
        {junkoxford}
\bibfield{author}{\bibinfo{person}{Vlad Howard Philip~N. Gallacher, John
  D.~Barash} {and} \bibinfo{person}{John Kelly}.} \bibinfo{year}{[n.
  d.]}\natexlab{}.
\newblock \bibinfo{title}{Junk News on Military Affairs and National Security:
  Social Media Disinformation Campaigns Against US Military Personnel and
  Veterans}.
\newblock   (\bibinfo{year}{[n. d.]}).
\newblock
\urldef\tempurl%
\url{http://comprop.oii.ox.ac.uk/publishing/working-papers/vetops/}
\showURL{%
\tempurl}


\bibitem[\protect\citeauthoryear{Ghosh, Viswanath, Kooti, Sharma, Korlam,
  Benevenuto, Ganguly, and Gummadi}{Ghosh et~al\mbox{.}}{2012}]%
        {ghosh2012understanding}
\bibfield{author}{\bibinfo{person}{Saptarshi Ghosh}, \bibinfo{person}{Bimal
  Viswanath}, \bibinfo{person}{Farshad Kooti}, \bibinfo{person}{Naveen~Kumar
  Sharma}, \bibinfo{person}{Gautam Korlam}, \bibinfo{person}{Fabricio
  Benevenuto}, \bibinfo{person}{Niloy Ganguly}, {and}
  \bibinfo{person}{Krishna~Phani Gummadi}.} \bibinfo{year}{2012}\natexlab{}.
\newblock \showarticletitle{Understanding and combating link farming in the
  twitter social network}. In \bibinfo{booktitle}{\emph{Proceedings of the 21st
  international conference on World Wide Web}}. ACM, \bibinfo{pages}{61--70}.
\newblock


\bibitem[\protect\citeauthoryear{Gilani, Farahbakhsh, Tyson, Wang, and
  Crowcroft}{Gilani et~al\mbox{.}}{2017a}]%
        {gilani2017bots}
\bibfield{author}{\bibinfo{person}{Zafar Gilani}, \bibinfo{person}{Reza
  Farahbakhsh}, \bibinfo{person}{Gareth Tyson}, \bibinfo{person}{Liang Wang},
  {and} \bibinfo{person}{Jon Crowcroft}.} \bibinfo{year}{2017}\natexlab{a}.
\newblock \showarticletitle{Of Bots and Humans (on Twitter)}. In
  \bibinfo{booktitle}{\emph{Proceedings of the 9th IEEE/ACM International
  Conference on Advances in Social Networks Analysis and Mining (ASONAM'17).
  https://doi. org/10.1145/3110025.3110090}}.
\newblock


\bibitem[\protect\citeauthoryear{Gilani, Kochmar, and Crowcroft}{Gilani
  et~al\mbox{.}}{2017b}]%
        {gilaniclassification}
\bibfield{author}{\bibinfo{person}{Zafar Gilani}, \bibinfo{person}{Ekaterina
  Kochmar}, {and} \bibinfo{person}{Jon Crowcroft}.}
  \bibinfo{year}{2017}\natexlab{b}.
\newblock \showarticletitle{Classification of Twitter Accounts into Automated
  Agents and Human Users}. In \bibinfo{booktitle}{\emph{Proceedings of the
  international conference on Advances in Social Network Analysis and Mining
  ASONAM}}.
\newblock


\bibitem[\protect\citeauthoryear{Han and Park}{Han and Park}{2013}]%
        {han2013efficient}
\bibfield{author}{\bibinfo{person}{Jin~Seop Han} {and}
  \bibinfo{person}{Byung~Joon Park}.} \bibinfo{year}{2013}\natexlab{}.
\newblock \showarticletitle{Efficient detection of content polluters in social
  networks}.
\newblock In \bibinfo{booktitle}{\emph{IT Convergence and Security 2012}}.
  \bibinfo{publisher}{Springer}, \bibinfo{pages}{991--996}.
\newblock


\bibitem[\protect\citeauthoryear{Hu, Tang, and Liu}{Hu et~al\mbox{.}}{2014}]%
        {hu2014online}
\bibfield{author}{\bibinfo{person}{Xia Hu}, \bibinfo{person}{Jiliang Tang},
  {and} \bibinfo{person}{Huan Liu}.} \bibinfo{year}{2014}\natexlab{}.
\newblock \showarticletitle{Online Social Spammer Detection.}. In
  \bibinfo{booktitle}{\emph{AAAI}}. \bibinfo{pages}{59--65}.
\newblock


\bibitem[\protect\citeauthoryear{Kayes and Iamnitchi}{Kayes and
  Iamnitchi}{2017}]%
        {kayes2017privacy}
\bibfield{author}{\bibinfo{person}{Imrul Kayes} {and} \bibinfo{person}{Adriana
  Iamnitchi}.} \bibinfo{year}{2017}\natexlab{}.
\newblock \showarticletitle{Privacy and security in online social networks: A
  survey}.
\newblock \bibinfo{journal}{\emph{Online Social Networks and Media}}
  \bibinfo{volume}{3} (\bibinfo{year}{2017}), \bibinfo{pages}{1--21}.
\newblock


\bibitem[\protect\citeauthoryear{Keller, Schoch, Stier, and Yang}{Keller
  et~al\mbox{.}}{2017}]%
        {keller2017manipulate}
\bibfield{author}{\bibinfo{person}{Franziska~B Keller}, \bibinfo{person}{David
  Schoch}, \bibinfo{person}{Sebastian Stier}, {and} \bibinfo{person}{JungHwan
  Yang}.} \bibinfo{year}{2017}\natexlab{}.
\newblock \showarticletitle{How to Manipulate Social Media: Analyzing Political
  Astroturfing Using Ground Truth Data from South Korea.}. In
  \bibinfo{booktitle}{\emph{ICWSM}}. \bibinfo{pages}{564--567}.
\newblock


\bibitem[\protect\citeauthoryear{Lee, Eoff, and Caverlee}{Lee
  et~al\mbox{.}}{2011}]%
        {lee2011seven}
\bibfield{author}{\bibinfo{person}{Kyumin Lee}, \bibinfo{person}{Brian~David
  Eoff}, {and} \bibinfo{person}{James Caverlee}.}
  \bibinfo{year}{2011}\natexlab{}.
\newblock \showarticletitle{Seven Months with the Devils: A Long-Term Study of
  Content Polluters on Twitter.}. In \bibinfo{booktitle}{\emph{ICWSM}}.
\newblock


\bibitem[\protect\citeauthoryear{Lee, Mahmud, Chen, Zhou, and Nichols}{Lee
  et~al\mbox{.}}{2014}]%
        {lee2014will}
\bibfield{author}{\bibinfo{person}{Kyumin Lee}, \bibinfo{person}{Jalal Mahmud},
  \bibinfo{person}{Jilin Chen}, \bibinfo{person}{Michelle Zhou}, {and}
  \bibinfo{person}{Jeffrey Nichols}.} \bibinfo{year}{2014}\natexlab{}.
\newblock \showarticletitle{Who will retweet this?: Automatically identifying
  and engaging strangers on twitter to spread information}. In
  \bibinfo{booktitle}{\emph{Proceedings of the 19th international conference on
  Intelligent User Interfaces}}. ACM, \bibinfo{pages}{247--256}.
\newblock


\bibitem[\protect\citeauthoryear{Muthiah, Huang, Arredondo, Mares, Getoor,
  Katz, and Ramakrishnan}{Muthiah et~al\mbox{.}}{2015}]%
        {muthiah2015planned}
\bibfield{author}{\bibinfo{person}{Sathappan Muthiah}, \bibinfo{person}{Bert
  Huang}, \bibinfo{person}{Jaime Arredondo}, \bibinfo{person}{David Mares},
  \bibinfo{person}{Lise Getoor}, \bibinfo{person}{Graham Katz}, {and}
  \bibinfo{person}{Naren Ramakrishnan}.} \bibinfo{year}{2015}\natexlab{}.
\newblock \showarticletitle{Planned Protest Modeling in News and Social
  Media.}. In \bibinfo{booktitle}{\emph{AAAI}}. \bibinfo{pages}{3920--3927}.
\newblock


\bibitem[\protect\citeauthoryear{Nasim, Charbey, Prieur, and Brandes}{Nasim
  et~al\mbox{.}}{2016a}]%
        {nasim2016investigating}
\bibfield{author}{\bibinfo{person}{Mehwish Nasim}, \bibinfo{person}{Rapha{\"e}l
  Charbey}, \bibinfo{person}{Christophe Prieur}, {and} \bibinfo{person}{Ulrik
  Brandes}.} \bibinfo{year}{2016}\natexlab{a}.
\newblock \showarticletitle{Investigating Link Inference in Partially
  Observable Networks: Friendship Ties and Interaction}.
\newblock \bibinfo{journal}{\emph{IEEE Transactions on Computational Social
  Systems}} \bibinfo{volume}{3}, \bibinfo{number}{3} (\bibinfo{year}{2016}),
  \bibinfo{pages}{113--119}.
\newblock


\bibitem[\protect\citeauthoryear{Nasim, Rextin, Khan, and Malik}{Nasim
  et~al\mbox{.}}{2016b}]%
        {nasim2016mobilehci}
\bibfield{author}{\bibinfo{person}{Mehwish Nasim}, \bibinfo{person}{Aimal
  Rextin}, \bibinfo{person}{Numair Khan}, {and}
  \bibinfo{person}{Muhammad~Muddassir Malik}.}
  \bibinfo{year}{2016}\natexlab{b}.
\newblock \showarticletitle{Understanding Call Logs of Smartphone Users for
  Making Future Calls}. In \bibinfo{booktitle}{\emph{18th International
  Conference on Human-Computer Interaction with Mobile Devices and Services}}.
  ACM.
\newblock


\bibitem[\protect\citeauthoryear{Nocaj, Ortmann, and Brandes}{Nocaj
  et~al\mbox{.}}{2014}]%
        {nocaj2014untangling}
\bibfield{author}{\bibinfo{person}{Arlind Nocaj}, \bibinfo{person}{Mark
  Ortmann}, {and} \bibinfo{person}{Ulrik Brandes}.}
  \bibinfo{year}{2014}\natexlab{}.
\newblock \showarticletitle{Untangling hairballs: From 3 to 14 degrees of
  separation}. In \bibinfo{booktitle}{\emph{22nd International Symposium, Graph
  Drawing 2014}}. \bibinfo{pages}{101--112}.
\newblock


\bibitem[\protect\citeauthoryear{Osborne, Lothian, Neale, Moscou, Nguyen, Chen,
  Kang, and Cooper}{Osborne et~al\mbox{.}}{2017}]%
        {grant}
\bibfield{author}{\bibinfo{person}{Grant Osborne}, \bibinfo{person}{Nick
  Lothian}, \bibinfo{person}{Grant Neale}, \bibinfo{person}{Terry Moscou},
  \bibinfo{person}{Andrew Nguyen}, \bibinfo{person}{Jie Chen},
  \bibinfo{person}{Wei Kang}, {and} \bibinfo{person}{Brenton Cooper}.}
  \bibinfo{year}{2017}\natexlab{}.
\newblock \showarticletitle{The beat the news system: Forecasting social
  disruption via modelling of online behaviours}.
\newblock \bibinfo{journal}{\emph{Journal of the Australian Institute of
  Professional Intelligence Officers}} (\bibinfo{year}{2017}).
\newblock


\bibitem[\protect\citeauthoryear{Ramakrishnan, Butler, Muthiah, Self, Khandpur,
  Saraf, Wang, Cadena, Vullikanti, Korkmaz, et~al\mbox{.}}{Ramakrishnan
  et~al\mbox{.}}{2014}]%
        {ramakrishnan2014beating}
\bibfield{author}{\bibinfo{person}{Naren Ramakrishnan},
  \bibinfo{person}{Patrick Butler}, \bibinfo{person}{Sathappan Muthiah},
  \bibinfo{person}{Nathan Self}, \bibinfo{person}{Rupinder Khandpur},
  \bibinfo{person}{Parang Saraf}, \bibinfo{person}{Wei Wang},
  \bibinfo{person}{Jose Cadena}, \bibinfo{person}{Anil Vullikanti},
  \bibinfo{person}{Gizem Korkmaz}, {et~al\mbox{.}}}
  \bibinfo{year}{2014}\natexlab{}.
\newblock \showarticletitle{'Beating the news' with EMBERS: forecasting civil
  unrest using open source indicators}. In
  \bibinfo{booktitle}{\emph{Proceedings of the 20th ACM SIGKDD international
  conference on Knowledge discovery and data mining}}. ACM,
  \bibinfo{pages}{1799--1808}.
\newblock


\bibitem[\protect\citeauthoryear{Rosen and Resnick}{Rosen and Resnick}{1980}]%
        {ranksize}
\bibfield{author}{\bibinfo{person}{Kenneth~T Rosen} {and}
  \bibinfo{person}{Mitchel Resnick}.} \bibinfo{year}{1980}\natexlab{}.
\newblock \showarticletitle{The size distribution of cities: an examination of
  the Pareto law and primacy}.
\newblock \bibinfo{journal}{\emph{Journal of Urban Economics}}
  \bibinfo{volume}{8}, \bibinfo{number}{2} (\bibinfo{year}{1980}),
  \bibinfo{pages}{165--186}.
\newblock


\bibitem[\protect\citeauthoryear{Saraf and Ramakrishnan}{Saraf and
  Ramakrishnan}{2016}]%
        {saraf2016embers}
\bibfield{author}{\bibinfo{person}{Parang Saraf} {and} \bibinfo{person}{Naren
  Ramakrishnan}.} \bibinfo{year}{2016}\natexlab{}.
\newblock \showarticletitle{EMBERS autogsr: Automated coding of civil unrest
  events}. In \bibinfo{booktitle}{\emph{Proceedings of the 22nd ACM SIGKDD
  International Conference on Knowledge Discovery and Data Mining}}. ACM,
  \bibinfo{pages}{599--608}.
\newblock


\bibitem[\protect\citeauthoryear{Su{\'a}rez-Serrato, Roberts, Davis, and
  Menczer}{Su{\'a}rez-Serrato et~al\mbox{.}}{2016}]%
        {suarez2016influence}
\bibfield{author}{\bibinfo{person}{Pablo Su{\'a}rez-Serrato},
  \bibinfo{person}{Margaret~E Roberts}, \bibinfo{person}{Clayton Davis}, {and}
  \bibinfo{person}{Filippo Menczer}.} \bibinfo{year}{2016}\natexlab{}.
\newblock \showarticletitle{On the influence of social bots in online
  protests}. In \bibinfo{booktitle}{\emph{International Conference on Social
  Informatics}}. Springer, \bibinfo{pages}{269--278}.
\newblock


\bibitem[\protect\citeauthoryear{Subrahmanian, Azaria, Durst, Kagan, Galstyan,
  Lerman, Zhu, Ferrara, Flammini, and Menczer}{Subrahmanian
  et~al\mbox{.}}{2016}]%
        {darpa2016}
\bibfield{author}{\bibinfo{person}{VS Subrahmanian}, \bibinfo{person}{Amos
  Azaria}, \bibinfo{person}{Skylar Durst}, \bibinfo{person}{Vadim Kagan},
  \bibinfo{person}{Aram Galstyan}, \bibinfo{person}{Kristina Lerman},
  \bibinfo{person}{Linhong Zhu}, \bibinfo{person}{Emilio Ferrara},
  \bibinfo{person}{Alessandro Flammini}, {and} \bibinfo{person}{Filippo
  Menczer}.} \bibinfo{year}{2016}\natexlab{}.
\newblock \showarticletitle{The DARPA Twitter bot challenge}.
\newblock \bibinfo{journal}{\emph{Computer}} \bibinfo{volume}{49},
  \bibinfo{number}{6} (\bibinfo{year}{2016}), \bibinfo{pages}{38--46}.
\newblock


\bibitem[\protect\citeauthoryear{Varol, Ferrara, Davis, Menczer, and
  Flammini}{Varol et~al\mbox{.}}{2017}]%
        {Varol2017}
\bibfield{author}{\bibinfo{person}{Onur Varol}, \bibinfo{person}{Emilio
  Ferrara}, \bibinfo{person}{Clayton~A Davis}, \bibinfo{person}{Filippo
  Menczer}, {and} \bibinfo{person}{Alessandro Flammini}.}
  \bibinfo{year}{2017}\natexlab{}.
\newblock \showarticletitle{{Online Human-Bot Interactions : Detection ,
  Estimation , and Characterization}}. In \bibinfo{booktitle}{\emph{Proceedings
  of the Eleventh International AAAI Conference on Web and Social Media (ICWSM
  2017) Online}}. \bibinfo{pages}{280--289}.
\newblock
\showISBNx{9781577357889}


\bibitem[\protect\citeauthoryear{Wang, Zubiaga, Liakata, and Procter}{Wang
  et~al\mbox{.}}{2015}]%
        {wang2015making}
\bibfield{author}{\bibinfo{person}{Bo Wang}, \bibinfo{person}{Arkaitz Zubiaga},
  \bibinfo{person}{Maria Liakata}, {and} \bibinfo{person}{Rob Procter}.}
  \bibinfo{year}{2015}\natexlab{}.
\newblock \showarticletitle{Making the most of tweet-inherent features for
  social spam detection on twitter}.
\newblock \bibinfo{journal}{\emph{arXiv preprint arXiv:1503.07405}}
  (\bibinfo{year}{2015}).
\newblock


\bibitem[\protect\citeauthoryear{Warzel and Loop}{Warzel and Loop}{[n. d.]}]%
        {buzzfeed}
\bibfield{author}{\bibinfo{person}{Charlie Warzel} {and} \bibinfo{person}{Emma
  Loop}.} \bibinfo{year}{[n. d.]}\natexlab{}.
\newblock \bibinfo{title}{Twitter Tells Congress It Found 200 Russian Accounts
  That Overlapped With Facebook}.
\newblock   (\bibinfo{year}{[n. d.]}).
\newblock
\urldef\tempurl%
\url{https://www.buzzfeed.com/charliewarzel/twitter-russian-accounts?utm_term=.immV81Pgd#.siA4vLxop}
\showURL{%
\tempurl}


\bibitem[\protect\citeauthoryear{Wu, Hu, Morstatter, and Liu}{Wu
  et~al\mbox{.}}{2017}]%
        {wu2017detecting}
\bibfield{author}{\bibinfo{person}{Liang Wu}, \bibinfo{person}{Xia Hu},
  \bibinfo{person}{Fred Morstatter}, {and} \bibinfo{person}{Huan Liu}.}
  \bibinfo{year}{2017}\natexlab{}.
\newblock \showarticletitle{Detecting Camouflaged Content Polluters.}. In
  \bibinfo{booktitle}{\emph{ICWSM}}. \bibinfo{pages}{696--699}.
\newblock


\bibitem[\protect\citeauthoryear{Yang, Harkreader, and Gu}{Yang
  et~al\mbox{.}}{2013}]%
        {yang2013empirical}
\bibfield{author}{\bibinfo{person}{Chao Yang}, \bibinfo{person}{Robert
  Harkreader}, {and} \bibinfo{person}{Guofei Gu}.}
  \bibinfo{year}{2013}\natexlab{}.
\newblock \showarticletitle{Empirical evaluation and new design for fighting
  evolving twitter spammers}.
\newblock \bibinfo{journal}{\emph{IEEE Transactions on Information Forensics
  and Security}} \bibinfo{volume}{8}, \bibinfo{number}{8}
  (\bibinfo{year}{2013}), \bibinfo{pages}{1280--1293}.
\newblock


\bibitem[\protect\citeauthoryear{Yang, Wilson, Wang, Gao, Zhao, and Dai}{Yang
  et~al\mbox{.}}{2014}]%
        {yang2014uncovering}
\bibfield{author}{\bibinfo{person}{Zhi Yang}, \bibinfo{person}{Christo Wilson},
  \bibinfo{person}{Xiao Wang}, \bibinfo{person}{Tingting Gao},
  \bibinfo{person}{Ben~Y Zhao}, {and} \bibinfo{person}{Yafei Dai}.}
  \bibinfo{year}{2014}\natexlab{}.
\newblock \showarticletitle{Uncovering social network sybils in the wild}.
\newblock \bibinfo{journal}{\emph{ACM Transactions on Knowledge Discovery from
  Data (TKDD)}} \bibinfo{volume}{8}, \bibinfo{number}{1}
  (\bibinfo{year}{2014}), \bibinfo{pages}{2}.
\newblock


\end{thebibliography}

\end{document}